\documentclass[12pt]{article}
\usepackage[utf8]{inputenc}
\usepackage[english]{babel}
\usepackage{amsmath}
\usepackage{mathrsfs}
\usepackage{amssymb}
\usepackage{amsthm}
\usepackage{amsfonts}
\usepackage{xspace}
\usepackage[normalem]{ulem}
\usepackage{graphicx}
\usepackage{multirow}
\usepackage{appendix}
\usepackage{enumerate}
\usepackage{natbib}
\usepackage{url} 
\usepackage{authblk}
\usepackage{float}
\usepackage{xcolor}
\usepackage{subcaption}

\usepackage{xcolor}
\usepackage{soul}
\usepackage{tikz}
\usetikzlibrary{calc}

\newtheorem{theorem}{Theorem}

\newtheorem{proposition}{Proposition}

\newtheorem{definition}{Definition}

\newcommand{\bbeta}{\boldsymbol{\beta}}
\newcommand{\bX}{\boldsymbol{X}}
\newcommand{\bQ}{\boldsymbol{Q}}
\newcommand{\bD}{\boldsymbol{D}}
\newcommand{\balpha}{\boldsymbol{\alpha}}

\begin{document}

\def\spacingset#1{\renewcommand{\baselinestretch}%
{#1}\small\normalsize}

\title{\bf Regularized regression on compositional trees with application to MRI analysis}

\newcommand*{\affaddr}[1]{#1} 
\newcommand*{\affmark}[1][*]{\textsuperscript{#1}}
\newcommand*{\email}[1]{\texttt{#1}}
\author{%
    Bingkai Wang\affmark[1], Brian S. Caffo\affmark[1], Xi Luo\affmark[2], Chin-Fu Liu\affmark[3], Andreia V. Faria\affmark[4],    Michael I. Miller\affmark[3], Yi Zhao\affmark[5],
    and for the Alzheimer's Disease Neuroimaging Initiative\footnote{Data used in preparation of this article were obtained from the Alzheimer's Disease Neuroimaging Initiative (ADNI) database (\protect\url{adni.loni.usc.edu}). As such, the investigators within the ADNI contributed to the design and implementation of ADNI and/or provided data but did not participate in analysis or writing of this report. A complete list of ADNI investigators can be found at: \protect\url{http://adni.loni.usc.edu/wp-content/uploads/how_to_apply/ADNI_Acknowledgement_List.pdf}} \\
    \affaddr{\affmark[1]\small Department of Biostatistics, Johns Hopkins Bloomberg School of Public Health} \\
    \affaddr{\affmark[2]\small Department of Biostatistics and Data Science,   The University of Texas 
Health Science Center at Houston} \\
    \affaddr{\affmark[3]\small Center for Imaging Science, Biomedical Engineering, Johns Hopkins University} \\
    \affaddr{\affmark[4]\small Department of Radiology, Johns Hopkins University School of Medicine} \\    
    \affaddr{\affmark[5]\small Department of Biostatistics, Indiana University School of Medicine} \\    
}

\date{\vspace{-5ex}}
\maketitle


\begin{abstract}
A compositional tree refers to a tree structure on a set of random variables where each random variable is a node and composition occurs at each non-leaf node of the tree. 
As a generalization of compositional data, compositional trees handle more complex relationships among random variables and appear in many disciplines, such as brain imaging, genomics and finance.
We consider the problem of sparse regression on data that are associated with a compositional tree and propose a transformation-free tree-based regularized regression method for component selection. 
The regularization penalty is designed based on the tree structure and encourages a sparse tree representation.
We prove that our proposed estimator for regression coefficients is both consistent and model selection consistent.
In the simulation study, our method shows higher accuracy than competing methods under different scenarios. 
By analyzing a brain imaging data set from studies of Alzheimer's disease, our method identifies meaningful associations between memory declination and volume of brain regions that are consistent with current understanding.
\end{abstract}

\noindent%

\spacingset{1.5} 
\newpage
\section{Introduction}
Compositional data refer to a type of data where data points are non-negative and the data vector of each subject or observational unit sums up to one. Compositional data appear in many disciplines, such as econometrics \citep{mullahy2015multivariate}, geology \citep{pawlowsky2006compositional} and epidemiology \citep{leite2016applying}. 

In the area of brain imaging, structural magnetic resonance imaging (MRI) and anatomical brain segmentation produce compositional data. For example, using 3-dimensional images acquired via structural MRI, a five-step brain segmentation introduced by \cite{mori2016mricloud} can partition the whole brain into regions at five granularity levels. At the most coarse level, the whole brain is segmented into telencephalon (left and right), diencephalon (left and right), metencephalon, mesencephalon and cerebrospinal fluid (CSF). At the finest level, the whole brain is segmented into 236 brain regions. The compositional data are then the fractional volumes of the 236 brain regions relative to the intracranial volume (ICV).

\begin{figure}[htbp]
    \centering
    \includegraphics[width=\textwidth]{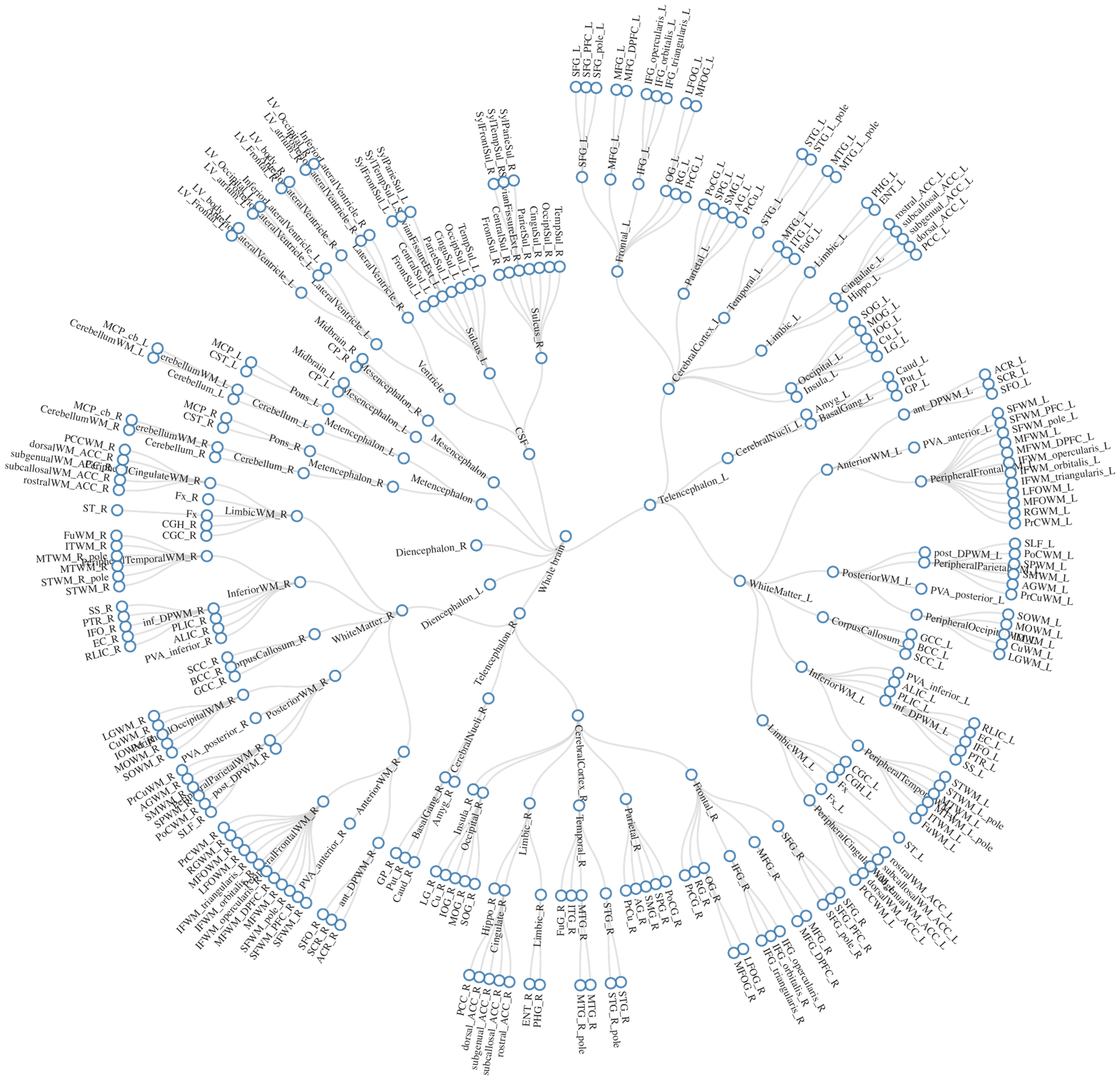}
    \vspace{-6cm}
    \caption{Compositional tree structure of the MRI data example. The tree is rooted at the whole brain. A brain region with suffix ``\_L''  or ``\_R'' indicates that the region is in the left or right hemisphere of the brain.}\label{supp-fig:tree-structure}
\end{figure}

In addition to composition, the volumetric data have a tree structure. 
In the first step of segmentation, the whole brain is partitioned into 7 brain regions. In the second step, each of the 7 brain regions created by the first step is further partitioned into smaller regions, which can be thought of as  tree branching.
Applied to all brain segmentation steps, this analogy makes a tree structure that is rooted at the whole brain and has 236 leaves, which are the brain regions at the finest segmentation. The tree structure is shown in Figure~\ref{supp-fig:tree-structure}.
A key feature of this tree structure is that the volume of a brain region is equal to the combined volume of its subregions (after one segmentation), which introduces extra composition among variables. We refer to this data structure as  ``compositional tree''. We note that the structure of compositional data is a special case of compositional trees, which only have leaves and a root.

Compositional trees appear in many disciplines. For example, \cite{wang2017structured} presented a compositional tree of microbiome data, where the compositional tree is formed by bacterial taxa at multiple taxonomic levels. 
Another example is the fractional market capitalization of stocks in the S\&P 500 index (relative to the total market capitalization of S\&P 500), where all 500 stocks are partitioned into 11 sectors and each sector is further broken down into industries according to the  Global Industry Classification Standard \citep{GICS}. The fractional market capitalization of a sector (or industry) is the summation of fractional market capitalization of stocks that are categorized into this sector (or industry).
Compared with compositional data, compositional trees provide more information about the relationships among variables and suggest grouping effects at different levels.

Although methods for analyzing compositional data or tree-structured data have been developed, little is known about how to deal with compositional trees. 
\cite{lin2014variable, fiksel2020transformation, ma2020quantile} studied regression methods for compositional data with or without regularization, but their results cannot be directly generalized to handle compositional trees. 
\cite{kim2012tree} proposed a tree lasso for estimating a sparse multi-response regression function, which did not consider compositional data.
To the best of our knowledge, the primary competitive work is \cite{wang2017structured}, which developed a tree-guided regularization method for structured sub-composition
selection. This work focused on a tree structure with composition on leaves, which is different from the compositional tree, where composition exists at each node of the tree. Furthermore, it did not handle boundary points (zeros or ones) in the data or cover asymptotic properties.

In this paper, we propose a regularized regression method to estimate the association between a dependent variable and independent variables that have a compositional tree structure. The regularization term is constructed from the tree structure, which is assumed to be known, and designed to achieve sparsity in both marginal and conditional effects from independent variables. Our model is transformation-free and able to handle boundary points (zeros or ones) in the data. We also establish consistency and model selection consistency of our estimators building on results from \cite{lee2015model}.

In the next section, we introduce an MRI data example. In Section~\ref{sec:model-assumptions}, we define the compositional tree and regression model. In Section~\ref{sec:estimation}, we present our proposed method to estimate the regression coefficients. We evaluate the performance of our proposed method through simulations in Section~\ref{sec:simulation}. The MRI data application is provided in Section~\ref{sec:data-analysis}. Section~\ref{sec:discussion} discusses future directions.

\section{Data example}\label{sec:data-example}
Data used in the preparation of this article were obtained from the Alzheimer’s Disease Neuroimaging Initiative (ADNI) database (\url{adni.loni.usc.edu}). The ADNI was launched in 2003 as a public-private partnership, led by Principal Investigator Michael W. Weiner, MD. The primary goal of ADNI has been to test whether serial magnetic resonance imaging (MRI), positron emission tomography (PET), other biological markers, and clinical and neuropsychological assessment can be combined to measure the progression of mild cognitive impairment (MCI) and early Alzheimer’s disease (AD).
We focus on the data set acquired by \cite{liu2019using} from the ADNI database.

The ADNI data set contains 819 subjects, which were diagnosed at the baseline as cognitive normal (229 subjects), mild cognitive impairment (MCI, 402 subjects) or Alzheimer's disease (AD, 188 subjects). For each subject, the composite memory score, MRI data, and a wide variety of demographic, behavioral and other non-imaging covariates were collected at several time points. We focus on the composite memory score and MRI data. The composite memory score was measured using data from the ADNI neuropsychological battery and validated by \cite{crane2012development}, with higher scores indicating better memory. The MRI data consist of high-resolution T1-weighted images, which are preprocessed and segmented through MRICloud (\url{www.MRICloud.org}, \citealp{mori2016mricloud}), a public platform for multi-contrast imaging segmentation and quantification. The preprocessing steps include orientation, inhomogeneity correction, and histogram matching following with large deformation diffeomorphic metric mapping (LDDMM).  Post to preprocessing,  segmentations were obtained by fusing the multi-atlas labelling method \citep{tang2013bayesian}.

For the MRI data, the five-level brain segmentation defines 321 brain regions, which form a tree structure.  
At the first level of brain segmentation, the whole brain is partitioned into 7 brain regions. At the second level of brain segmentation, each of the 7 brain regions is further segmented into smaller regions. At the finest level, there are 236 brain regions. Figure~\ref{supp-fig:tree-structure} displays the tree structure of the 236 brain regions at the finest granularity. For each brain region, we extracted its volume. 
Based on the five-level brain segmentation procedure, the volume of a brain region is equal to the combined volume of its subregions (after one segmentation). Furthermore, the combined volume of brain regions at the finest level is equal to the ICV. 

Structural MRI data have been commonly used to identify biomarkers of AD \citep{vemuri2010role}. For example, the density of neurofibrillary tangles is an established pathological hallmark of AD, which can be reflected by MRI.
We hence focus on the association between memory declination, a common symptom of AD, and brain volumes. For each subject, we use the MRI data acquired at the initial screening, i.e. first time point, and the composite memory score acquired on the same day as the MRI scan or the first post-imaging measurement.


\section{Model and Assumptions}\label{sec:model-assumptions}
\subsection{Compositional tree}
We first define tree structure using notation from graph theory. 
Let $V = \{X_1, \dots, X_p\}$ be a set of random variables with $0\leq X_j \leq 1$ for $j = 1,\dots,p$. Let $E$ be a set of directed edges among $X_1, \dots, X_p$ with $E \subset \{(X_j \rightarrow X_k): X_j, X_k \in V\}$. For each edge $(X_j \rightarrow X_k) \in E$, we call $X_j$ the parent of $X_k$, and $X_k$ the child of $X_j$. 
$X_j$ is a leaf node if it has no child and a root node if it has no parent. $X_j$ is an ancestor of $X_k$ if the directed edges in $E$ can form a directed path from $X_j$ to $X_k$, for example, $(X_j \rightarrow X_s), (X_s \rightarrow X_k) \in E$. 

\begin{definition}\label{def:hierarchical-tree}
$(V, E)$ forms a tree if (1) no $X_j$ is an ancestor of itself (i.e., $E$ not containing any directed cycle), (2) $V$ contains only one root node and (3) each $X_j$ has at most one parent.
\end{definition}

\begin{figure}[htbp]
\centering
\begin{tikzpicture}
\def\aleft[slope=#1,height=#2] (#3) to (#4) node #5; {
    \node (#4) at ($(#3)+({180+#1}:{#2/cos(90-#1})$) {#5};
    \draw[thick, <-] (#4.north)--(#3.south);
}
\def\aright[slope=#1,height=#2] (#3) to (#4) node #5; {
    \node (#4) at ($(#3)+({-#1}:{#2/cos(90-#1})$) {#5};
    \draw[thick, <-] (#4.north)--(#3.south);
}
\def\binary[slope left=#1,slope right=#2,left height=#3,right height=#4] (#5) to (#6) node #7 and (#8) node #9; {
    \node (#6) at ($(#5)+({180+#1}:{#3/cos(90-(#1))})$) {#7};
    \node (#8) at ($(#5)+({-#2}:{#4/cos(90-(#2))})$) {#9};
    \draw[thick, <-] (#6.north)--(#5.south);
    \draw[thick, <-] (#8.north)--(#5.south);
}
\node (root) at (0,0) {$X_{10}$};
\aleft[slope=50,height=1.5] (root) to (9) node {$X_9$};
\aright[slope=50,height=3] (root) to (8) node {$X_8$};
\aleft[slope=55,height=3] (9) to (1) node {$X_1$};
\aright[slope=60,height=1.5] (9) to (7) node {$X_7$};
\aleft[slope=60,height=1.5] (7) to (2) node {$X_2$};
\aright[slope=60,height=1.5] (7) to (3) node {$X_3$};
\aleft[slope=60,height=1.5] (8) to (4) node {$X_4$};
\aleft[slope=90,height=1.5] (8) to (5) node {$X_5$};
\aright[slope=60,height=1.5] (8) to (6) node {$X_6$};
\end{tikzpicture}
\caption{An example of a tree with $p = 10$.}\label{fig:example-ht}
\end{figure}
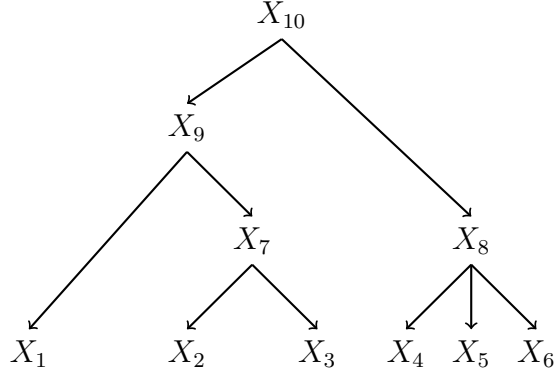

In Definition~\ref{def:hierarchical-tree}, condition (1) defines a directed acyclic graph, and  conditions (2) and (3) are often made in defining a rooted tree in graph theory. 
Figure~\ref{fig:example-ht} gives an example of tree structure with $V = \{X_1,\dots, X_{10}\}$ and $E = \{(X_{10} \rightarrow X_9), (X_{10} \rightarrow X_8), (X_9 \rightarrow X_1), (X_9 \rightarrow X_7), (X_7 \rightarrow X_2), (X_7 \rightarrow X_3), (X_8 \rightarrow X_4), (X_8 \rightarrow X_5), (X_8 \rightarrow X_6)\}$.

In our data example, we can define a tree given the hierarchical brain segmentation. Let each $X_j, j = 1,\dots, 321$, represent the volume of a brain region $j$ and let $V$ be the set of all $X_j$.
We regard $X_j$ as the parent of $X_k$ if brain region $k$ is a subregion of $j$ defined by one-step segmentation (i.e. there is no other subregion  of $j$ that contains $k$). If $X_j$ is the parent of $X_k$, we also call brain region $j$ the parent of brain region $k$. Then the edge set $E$ is defined as the collection of all parent-child relationships among brain regions and the (only) root node is the ICV.
For the $(V,E)$ defined above, condition (1) of Definition~\ref{def:hierarchical-tree} holds by construction, condition (2) follows because the root node is the ICV, and condition (3) results from the fact that a region cannot be part of two disjoint bigger regions.


Although we define the tree structure using notations of graph theory, we emphasize that we do not associate the tree structure with conditional independence or causal diagrams, such as in graphical probabilistic models \citep{pearl2009causality}. Our tree solely represents the hierarchical structure among $X_1, \dots, X_p$ and is used to add compositional constraints, as described below.
Our goal is to study the association between an outcome of interest and covariates $X_j, j = 1,\dots, p$, instead of the relationships among covariates.

Consider compositional constraints on $(X_1,\dots, X_p)$ complying with the tree structure. 
Denoting $q$ as the number of leaf nodes, we can arrange the indices of $X_1, \dots, X_p$ such that the first $q$ variables $(X_1, \dots, X_q)$ are the leaf nodes. For each $j = 1, \dots, p$, let $c(j) = \{k: (X_j \rightarrow X_k) \in E\}$ denote the index set of children of $X_j$ and let $|c(j)|$ denote the cardinality of $c(j)$ (i.e., the number of children of $X_j$). We then have the following definition of a compositional tree.

\begin{definition}\label{def:composition-ht}
Assume $(V, E)$ forms a tree and $X_1, \dots, X_q$ are the leaf nodes.
Then $(V, E)$ forms a compositional tree if (1) $\sum_{j=1}^q X_j = 1$ and
(2) $X_j = \sum_{k \in c(j)} X_k$ for each $j > q$.
\end{definition}

In Definition~\ref{def:composition-ht}, condition (1) imposes a compositional constraint on the leaf nodes. 
Condition (2) requires that each parent node is equal to the summation of its children. 
Conditions (1) and (2) together imply that the root node is a constant 1.
In the example shown in Figure~\ref{fig:example-ht}, the constraints for a compositional tree are $X_7 = X_2 + X_3$, $X_8 = X_4 + X_5 + X_6$, $X_9 = X_1 + X_7$ and $X_{10}=X_{8}+X_{9}=1$.
For the case that $X_j$ has only one child $X_k$, Definition~\ref{def:composition-ht} implies that $X_j = X_k$ and we hence drop $X_k$ to avoid any replicate.
In this paper, we assume the compositional tree $(V,E)$ for a column vector of random variables $\bX = (X_1,\dots, X_p)^t$ is known. 

Compositional trees generalize the structure on compositional data by allowing more constraints on $\bX$ and are less studied.
Although $X_{q+1}, \dots, X_p$ are linear combinations of leaf nodes, they still provide information on the structure of $\bX$ and can help interpret conditional effects (defined in Section~\ref{subsec:model} below).
To simplify notation, we say that $\bX$ has a compositional tree structure if the associated $(V,E)$ forms a compositional tree.


In our data example, brain regions $\bX$ defined above have a compositional tree structure. Since the summation of all leaf node volumes is the ICV, then condition (1) of Definition~\ref{def:composition-ht} requires that the volumetric data are normalized by the ICV such that each person has a total brain volume 1. This is common practice in MRI analysis, since the  ICV is typically only meaningfully related to physical size. Alternative strategies remove ventricular volumes and then study regional volumes relative to total brain volume (i.e., studying tissue composition).
We include the ventricular volumes and normalize by ICV, since they are an important aspect of understanding progressive tissue loss in a disorder like AD.
Condition (2) of Definition~\ref{def:composition-ht} states that the volume of each brain region is equal to the combined volume of all its children, which follows by definition that each brain region is partitioned into subregions with no volume left undefined.

When dealing with compositional data, most current models work on a transformed space, for example, Isometric logratio transformations \citep{egozcue2003isometric} or log ratio transformations \citep{papke1996econometric}. Although such transformations provide convenience in estimation, they cannot handle boundary values in $\bX$ and add difficulty to interpretation \citep{fiksel2020transformation}. We hence work on the original space $\{\bX : X_j \ge 0, j =1,\dots,p; \sum_{j=1}^p X_j = 1; X_j = \sum_{k \in c(j)} X_k, j  = q+1,\dots, p\}$. 

For a compositional tree, the vector space spanned by $\bX$ has dimension (at most) $q < p$, which causes rank deficiency in many regression models. 
An alternative way is to model $X_j = \sum_{k \in c(j)} X_k + \varepsilon_j$, where $\varepsilon_j$ is an independent Gaussian noise, following the method of \cite{shojaie2010penalized}.
Although this method does not have the competing issue of rank deficiency, as long as the covariance matrix of $(\varepsilon_1, \dots, \varepsilon_p)$ is positive definite, we have $\varepsilon_j = 0$ almost always, which violates Gaussian modeling assumptions.

\subsection{Linear model, parameter identifiability and interpretation}\label{subsec:model}

Let $Y$ be the outcome of interest. We assume the following linear model
\begin{equation}\label{eq:linear-model}
    Y = \sum_{j=1}^p \beta_j X_j + \varepsilon= \bbeta^\top \bX + \varepsilon,
\end{equation}
where $\bbeta = (\beta_1, \dots, \beta_p)^\top$ is a column vector of unknown parameters, $\bX$ has a compositional tree structure with $q$ leaf nodes, and $\varepsilon \sim N(0,\sigma^2)$ is independent of $\bX$. 
Since the root node of a compositional tree is a constant 1 and included in $\bX$, the intercept term is omitted from model  (\ref{eq:linear-model}).
For $i = 1,\dots, n$, let $(\bX_i, \varepsilon_i)$ be independent, identically distributed samples from the joint distribution of $(\bX, \varepsilon)$ and let $Y_i = \bbeta^\top \bX_i + \varepsilon_i$.

Since $\bX$ is rank deficient (with rank at most $q$), $\bbeta$ is not unique. 
Due to this fact, each $\beta_j , j = 1,\dots, p$ is not interpretable without further assumptions. To overcome this difficulty, we impose the following $p-q$ linear constraints on $\bbeta$:
\begin{equation} \label{eq:beta-constraint}
\sum_{k \in c(j)} \beta_k  = 0 \textrm{ for all } j > q,
\end{equation}
which uniquely define a $\bbeta \in \mathcal{S}$ (as shown in the Supplementary Material).
Linear constraints~(\ref{eq:beta-constraint}) require that, for each $X_j$ that is not a leaf node, the average effect of its children on $Y$ is $0$.
Then, each $\beta_k$ can be interpreted as the deviation effect of $X_k$ from the effect of its parent, $X_j$, on $Y$.
To show this, consider the following derivation using Definition~\ref{def:composition-ht}:
\begin{displaymath}
\beta_jX_j + \sum_{k \in c(j)} \beta_k X_k = \beta_j = \left(\beta_j + \frac{1}{|c(j)|} \sum_{l \in c(j)} \beta_l\right) X_j + \sum_{k \in c(j)}\left(\beta_k - \frac{1}{|c(j)|} \sum_{l \in c(j)} \beta_l\right) X_k,
\end{displaymath}
which implies that the average coefficient of children of $X_j$ can be absorbed into the coefficient of $X_j$ and hence the remaining coefficients of $X_k, k \in c(j)$ are the deviations from $X_j$. By repeating this procedure recursively from leaf nodes to the root node, we get all coefficients satisfying linear constraints (\ref{eq:beta-constraint}) with the desired interpretation. For conciseness, $\beta_k$ is referred to as the ``conditional deviation effect'' throughout, since its interpretation is conditioning on the parent of $X_k$, i.e. the parent of $X_k$ held constant.
Let $X_p$ denote the root node and $a(j)$ be the index set of ancestors of $X_j$. Then the linear model~(\ref{eq:linear-model}) with constraints~(\ref{eq:beta-constraint}) can be formulated as:
\begin{align}
     Y &= \beta_p + \sum_{j=1}^q \alpha_j X_j + \varepsilon = \beta_p + \balpha^\top \bX_{leaf} + \varepsilon,\label{eq:linear-model-2}\\
     &\quad \textrm{subject to} \quad \sum_{j=1}^q \alpha_j = 0, \label{eq:alpha-constraint}
\end{align}
where $\beta_{p}$ is the regression coefficient of the root node $X_p$ and serves as the intercept, $\balpha = (\alpha_1,\dots, \alpha_q)^\top$ with $\alpha_j = \beta_j + \sum_{k \in a(j)\setminus\{p\}} \beta_k$ and $\bX_{leaf} = (X_1, \dots, X_q)^\top$ is the vector of leaf nodes. 
We assume that the only linear constraint on $\bX_{leaf}$ is $\sum_{j=1}^q X_j = 1$, i.e., no component of $\bX_{leaf}$ being a linear combination of the others.
The model~(\ref{eq:linear-model-2}) not only provides direct interpretation of marginal associations between $Y$ and $\bX_{leaf}$ (which we introduce below), but is also useful for estimating $\bbeta$ in Section~\ref{sec:estimation}.

Compared to model~(\ref{eq:linear-model}), model~(\ref{eq:linear-model-2}) only uses the leaf nodes. Each $\alpha_j$ is the aggregation of conditional deviation effect of ancestors of $X_j$ excluding the root node. 
For each $j = 1,\dots, q$, $\alpha_j$ can be interpreted as the deviation effect of $X_j$ from the average effect of all leaf nodes, referred to as the ``marginal deviation effect'' throughout for conciseness. If $X_j$ is increased by $\delta$ at the expense of another leaf node, $X_k$, i.e., $X_k$ decreased by $\delta$, then $Y$ is changed by $\alpha_j - \alpha_k$.
If $X_j$ is increased by $\delta$ at the expense of all other leaf nodes evenly, i.e., $X_k$ decreased by $\delta/(q-1)$ for all $k \le q$ and $k \ne j$, then $Y$ is changed by $\frac{q}{q-1} \delta \alpha_j$.
Without the constraint~(\ref{eq:alpha-constraint}), $\alpha_j$ is not identifiable, since $\sum_{j=1}^q X_j =1$. However, $\beta_p + \alpha_j$ would still be identifiable and a fact that we use for estimation in Section~\ref{sec:estimation}. This statement is formally described in Proposition~\ref{prop:identifiability} below, which is proven in the Supplementary Material. 

\begin{proposition}\label{prop:identifiability}
Assume that $\sum_{j=1}^q X_j =1$ and no component of $\bX_{leaf}$ is a linear combination of the others. If two sets of parameters $(\beta_p, \alpha_1, \dots, \alpha_q)$ and $(\overline\beta_p, \overline\alpha_1, \dots, \overline\alpha_q)$ both satisfy model~(\ref{eq:linear-model-2}), then $\beta_p + \alpha_j = \overline\beta_p +  \overline\alpha_j$ for each $j = 1,\dots, q$.
\end{proposition}


Compared to $\balpha$, which is the marginal deviation effect, $\bbeta$ is the conditional deviation effect, offering flexibility for interpreting various conditional effects. For example, in the tree structure shown in Figure~\ref{fig:example-ht}, $\beta_2 + \beta_7$ represents the deviation effect of $X_2$ from $X_9$, and $\beta_7 + \beta_9$ represents the marginal deviation effect of $X_7$ (i.e. conditioning on a constant $X_{10}$).

In our data example, both $\balpha$ and $\bbeta$ are scientifically meaningful. The marginal deviation effect represents the effect of the fractional volume of a leaf region on memory, while the conditional deviation effect is the residual effect of the fractional volume of a brain region on memory after removing the effect of its ancestors on memory.


\section{Estimation}\label{sec:estimation}
Let $\bbeta^*$, $\balpha^*$ denote the true parameters that satisfy model~(\ref{eq:linear-model}) with constraints~(\ref{eq:beta-constraint}) and model~(\ref{eq:linear-model-2}) with constraint~(\ref{eq:alpha-constraint}) respectively.
Our goal is to estimate $\balpha^*$ and $\bbeta^*$. 
Since $p$ and $q$ are potentially large ($p = 321$ and $q=236$ for our data example), we propose a new regularization term  based on lasso for component selection \citep{lin2014variable} and fused lasso \citep{tibshirani2005sparsity} and perform regularized regression to achieve sparsity in both $\widehat{\balpha}$ and $\widehat{\bbeta}$.  In the method described below, we first estimate $\balpha^*$ using the generalized lasso \cite{tibshirani2011solution} and then calculate $\widehat{\bbeta}$ based on $\widehat{\balpha}$ by solving linear systems. 

\subsection{Regularization}
For any $\bbeta \in \mathbb{R}^p$ and $\balpha \in \mathbb{R}^q$, consider the regularization term $$P(\balpha, \bbeta, \eta)  = \eta P_1(\balpha) + (1-\eta) P_2(\bbeta),$$ where $\eta \in [0,1]$ is a tuning parameter adjusting the weight between two $P_1(\balpha)$ and $P_2(\bbeta)$,
\begin{align*}
    P_1(\balpha) &= \sum_{j=1}^q \Big|\alpha_j - \frac{1}{q}\sum_{k=1}^q\alpha_k\Big|, \\
    P_2(\bbeta) &=  \sum_{j = q+1}^p \sum_{s=1}^{|c(j)|-1}|\beta_{j_s} - \beta_{j_{s+1}}|,
\end{align*}
where $c(j)$ is the index set of children of $X_j$ with the elements in $c(j)$ encoded as $j_1,\dots, j_{|c(j)|}$. 
$P_1(\balpha)$ selects leaf nodes with non-zero marginal deviation effects. If $\alpha_j=\frac{1}{q}\sum_{k=1}^q\alpha_k$, then changing $X_j$ at the expense of all other leaf nodes  evenly will not result in changes of $Y$. This penalty is known as the lasso for component selection, which is also seen in \cite{wang2017structured} for dealing with compositional data.
In $P_2(\bbeta)$, for each $X_j$ with $j > q$, we penalize the difference among coefficients of its children using the fused lasso penalty. If $|\beta_{j_s} - \beta_{j_{s+1}}| = 0$ for all $s = 1,\dots, |c(j)|-1$, which means all children of $X_j$ have no conditional deviation effect, then the component $\sum_{k \in c(j)} \beta_k X_k=\beta_{j_1} X_j$, resulting in a sparse representation of linear model~(\ref{eq:linear-model}). Combined with the linear constraints~(\ref{eq:beta-constraint}), the above case is also equivalent to $\beta_k =0$ for all $k \in c(j)$.
The following proposition gives some properties of $P(\balpha, \bbeta, \eta)$.
\begin{proposition}\label{prop:penalty}
Given linear constraints~(\ref{eq:beta-constraint}), there exists a matrix $\bD(\eta) \in \mathbb{R}^{(2q-1)\times q}$ such that $P(\balpha, \bbeta, \eta) = ||\bD(\eta)\balpha||_1$ and $\bD(\eta)\boldsymbol{1}_q = \boldsymbol{0}_q$, where $||\cdot||_1$ is the $L_1$-norm,  $\boldsymbol{1}_q, \boldsymbol{0}_q \in \mathbb{R}^q$ are column vectors with all entries 1, 0 respectively.
\end{proposition}
Proposition~\ref{prop:penalty} implies that the penalty $P(\balpha, \bbeta, \eta)$ can be formulated as a function of $\balpha$ and $\eta$, making it possible to perform regularized regression based on model~(\ref{eq:linear-model-2}), which does not involve $\bbeta$. Furthermore, this penalty is invariant with respect to constant change of $\balpha$ (i.e., $||\bD(\eta)\balpha||_1 = ||\bD(\eta)(\balpha + C\boldsymbol{1}_q)||_1$ for any $C \in \mathbb{R}$), which makes it equivalent to penalize on $\balpha + \beta_q \boldsymbol{1}_q$ as we do below.
We prove Proposition~\ref{prop:penalty} and show how $\bD(\eta)$ is constructed in the Supplementary Material.
\subsection{Estimating $\balpha^*$}
We estimate $\balpha^*$ by $\widehat{\balpha} = \widehat{\widetilde{\balpha}} -  \boldsymbol{1}_q\boldsymbol{1}_q^\top\widehat{\widetilde{\balpha}}$, where
\begin{equation}\label{eq:estimate-alpha}
    \widehat{\widetilde{\balpha}}= \arg\min_{\widetilde{\balpha}} \frac{1}{n}\sum_{i=1}^n\left(Y_i - \widetilde{\balpha}^\top\bX_{leaf,i}\right)^2  + \lambda||\bD(\eta)\widetilde{\balpha}||_1
\end{equation}
with $\bX_{leaf,i} = (X_{i1},\dots, X_{iq})^\top$, $\widetilde{\balpha} = \balpha + \beta_p\boldsymbol{1}_q$ and $\lambda > 0$ being the tuning parameter. In equation~(\ref{eq:estimate-alpha}), $\widehat{\widetilde{\balpha}}$ is an estimate of $\widetilde{\balpha}$, which is identifiable as discussed in Section~\ref{subsec:model} and does not involve any linear constraints. Then, $\widehat{\balpha}$ is constructed by imposing the constraint~(\ref{eq:alpha-constraint}), i.e. centering $\widehat{\widetilde{\balpha}}$. We note that the regularization term $\lambda||\bD(\eta)\widetilde{\balpha}||_1$ imposes the desired sparsity on $\balpha$, since $\lambda||\bD(\eta)\widetilde{\balpha}||_1 = \lambda||\bD(\eta){\balpha}||_1$ given Proposition~\ref{prop:penalty}. 

For any $\balpha \in \mathbb{R}^q$ and given $\eta \in [0,1]$, let $\mathcal{S}(\balpha)$ be the support of $\bD(\eta)\balpha$, i.e., $\mathcal{S}(\balpha) = \{j \in \{1,\dots, 2q-1\}: \boldsymbol{e}_j^\top \bD(\eta)\balpha \ne 0\}$ with $\boldsymbol{e}_j \in \mathbb{R}^{2q-1}$ being a column vector with the $j$-th entry 1 and the rest 0. Let $\mathcal{M} = \{\balpha: \mathcal{S}(\balpha) \subset \mathcal{S}(\balpha^*)\}$ denote the model subspace of interest. That is, for $\balpha \in \mathcal{M}$, an entry of $\bD(\eta)\balpha$ is non-zero only if the corresponding entry of $\bD(\eta)\balpha^*$ is non-zero.
The following theorem gives consistency and model selection consistency of $\widehat{\balpha}$, which is adapted from Corollary 4.2 of \cite{lee2015model}.
\begin{theorem}\label{thm1}
Given $\eta \in [0,1]$, we assume $\{\bX_{leaf,i}\}_{i=1}^n$ satisfies restricted strong convexity (RSC) on $\mathcal{M}$ and irrepresentability, which we define in the Supplementary Material. For $\lambda = C_1 \sigma\sqrt{\frac{\log q}{n}}$, $\widehat{\balpha}$ is unique and, with probability at least $1-2/q$,
\begin{enumerate}
    \item (consistency) $||\widehat{\balpha} - \balpha^*||_2 \le C_2 \sigma\sqrt{\frac{\log q}{n}}$,
    \item (model selection consistency) $\widehat{\balpha} \in \mathcal{M}$,
\end{enumerate}
where $||\cdot||_2$ is the $L_2$-norm and $C_1, C_2$ are known constants given in the Supplementary Material. 
\end{theorem}
Theorem~\ref{thm1} implies that when $q$ and $n/\log(q)$ are large, then, with high probability, our estimate $\widehat{\balpha}$ is close to the truth and does not contain false positives (non-zero effect of inactive predictors with respect to $\bD(\eta)$). The RSC assumption is typically satisfied when $X_{leaf,i}$ follows a multivariate normal distribution \citep{raskutti2010restricted}. The irrepresentability assumption requires that the active predictors (with respect to $\bD(\eta)$) are not overly well-aligned with the inactive predictors. This is achieved when the inactive predictors are orthogonal or nearly-orthogonal to the active predictors \citep{lee2015model}. We provide a detailed description and discussion of these assumptions in the Supplementary Material.

Given $\eta$, the optimization problem~(\ref{eq:estimate-alpha}) can be solved by the \textit{genlasso} package \citep{tibshirani2011solution} in R software. To select the tuning parameter $\lambda$, we propose to use the Akaike information criterion (AIC, \citealp{akaike1998selected}) or Bayesian information criterion (BIC, \citealp{schwarz1978estimating}). Let
\begin{displaymath}
IC_{\gamma}(\eta, \lambda) = n \log\left\{\sum_{i=1}^n\left(Y_i - \widehat{\widetilde{\balpha}}^\top\bX_{leaf,i}\right)^2\right\} + \gamma\ \textrm{df}(\eta, \lambda),
\end{displaymath}
where $\gamma$ is a complexity factor, $\textrm{df}(\eta, \lambda)$ is the effective number of parameters in $\widehat{\widetilde{\balpha}}$. $IC_{\gamma}(\eta, \lambda)$ refers to AIC if $\gamma = 2$ and BIC if $\gamma = \log(n)$.
For any $\eta \in [0,1]$, define $\widehat{\lambda}(\eta) = \arg\min_{\lambda\ge 0}IC_{\gamma}(\eta, \lambda)$. We select the tuning parameters $\widehat{\eta} = \arg\min_{\eta \in [0,1]} IC_{\gamma}(\eta, \widehat{\lambda}(\eta))$ and $\widehat{\lambda} = \widehat{\lambda}(\widehat{\eta})$. An alternative method to tune parameters is cross-validation, but we do not consider it here, since it would dramatically increase the computation complexity and performs similarly to AIC.

\subsection{Estimating $\bbeta^*$}
Given $\widehat{\balpha}$, we calculate $\widehat{\bbeta}$ as follows. 
Since $\alpha_j = \beta_j + \sum_{k \in a(j)} \beta_k - \beta_p$ for $j = 1,\dots, q$, we can construct a matrix $\bQ_1 \in \mathbb{R}^{q\times p} $ such that $\bQ_1 \bbeta = \balpha$.
Since $\bbeta$ also satisfies linear constraints~(\ref{eq:beta-constraint}), we can construct another matrix $\bQ_2 \in \mathbb{R}^{(q-p)\times p}$ such that $\bQ_2 \bbeta = \boldsymbol{0}_{p-q}$.
Denoting $\bQ = (\bQ_1^\top, \bQ_2^\top)^\top$, then $\widehat{\bbeta}$ is calculated by solving the linear system
\begin{equation}\label{eq:estimate-beta}
\bQ\bbeta = \left(\begin{array}{c}
    \widehat{\balpha}  \\
    \boldsymbol{0}_{p-q} 
\end{array}\right).
\end{equation}
The following theorem implies that $\widehat{\bbeta}$ is uniquely determined by $\widehat{\balpha}$ (i.e. $\bQ$ is invertible) and is consistent and model selection consistent under the same conditions as $\widehat{\balpha}$.
\begin{theorem}\label{thm:beta}
Let $C_1, C_2, \lambda$ and $\mathcal{M}$ be the quantities defined in Theorem~\ref{thm1}. Given the same assumptions made in Theorem~\ref{thm1}, $\widehat{\bbeta}$ is uniquely determined by $\widehat{\balpha}$ and, with probability at least $1-2/q$,
\begin{enumerate}
    \item (consistency) $||\widehat{\bbeta} - \bbeta^*||_2 \le ||\bQ^{-1}||_2C_2 \sigma\sqrt{\frac{\log q}{n}}$,
    \item (model selection consistency) $\bQ_1\widehat{\bbeta} \in \mathcal{M}$.
\end{enumerate}
\end{theorem}

An alternative method to estimate $\bbeta$ is solving a constrained optimization problem following \cite{lin2014variable}:
\begin{align*}
 \widehat\bbeta  =& \arg \min_{\bbeta} \frac{1}{n}\sum_{i=1}^n\left(Y_i - \bbeta^\top\bX_i\right)^2  + \lambda P(\alpha, \bbeta, \eta),\\
   \textrm{subject to}\quad & \ \textrm{linear constraints~(\ref{eq:beta-constraint})}.
\end{align*}
However, this method has to handle the rank deficiency of $\bX$ and $p-q$ linear constraints. 
If $q$ is much smaller than $p$, then the linear constraints can be large, which may cause bias and increased computational complexity. 
In our proposed method, these two issues are avoided by using two steps to estimate $\bbeta^*$ (first estimating $\balpha^*$ and then $\bbeta^*$).

To the best of our knowledge, we are the first to study regularized regression on a compositional tree and provide consistency and model selection consistency. \cite{kim2012tree} developed a tree-guided group lasso method, but their goal was to analyze multi-response data and they did not consider composition. \cite{lin2014variable} used the lasso for component selection in compositional data and their optimization problem is a special case of ours, setting $\eta = 1$. \cite{wang2017structured} proposed TASSO to perform penalized regression on compositional data with a hierarchical tree structure, but they did not consider compositional trees or provide asymptotic results. We compare our method with lasso for component selection and TASSO in both simulations and MRI data application below.

\section{Simulation study}\label{sec:simulation}
In this simulation study, we consider 4 data generating distributions, which cover combinations of the following settings: a binary compositional tree or the MRI-motivated compositional tree, and leaf or stem effects. A binary compositional tree is a compositional tree where each parent has two children, while the MRI-motivated compositional tree represents the same tree structure as our data example (where a parent node may have more than two children). Leaf effects stand for linear models where the true effects (non-zero $\beta_j$) are only from nodes near the leaves, while stem effects mean that true effects are only from nodes near the root. Different from the leaf effects where both $\balpha^*$ and $\bbeta^*$ are sparse, stem effects will lead to non-sparse $\balpha^*$.

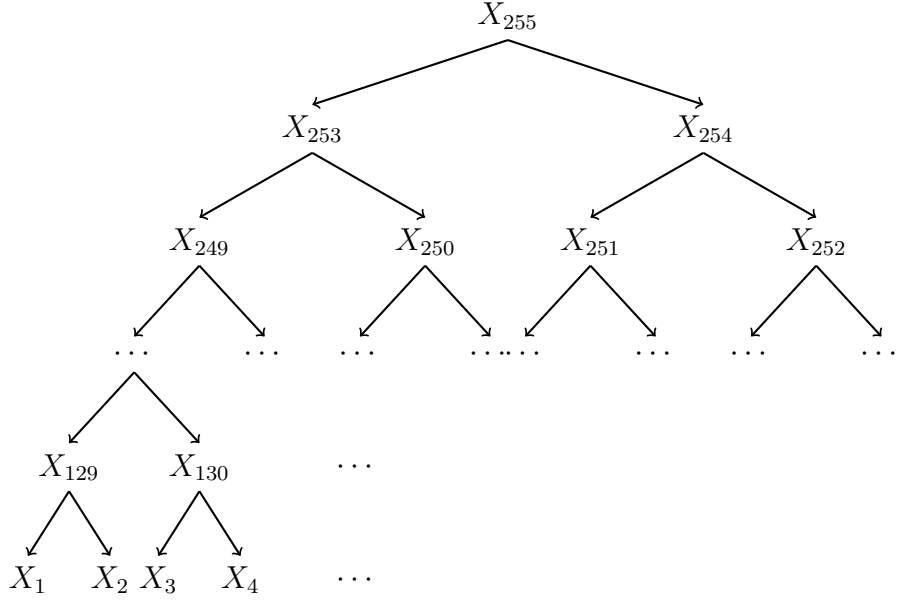
\begin{figure}[ht]
\centering
\begin{tikzpicture}
\def\aleft[slope=#1,height=#2] (#3) to (#4) node #5; {
    \node (#4) at ($(#3)+({180+#1}:{#2/cos(90-#1})$) {#5};
    \draw[thick, <-] (#4.north)--(#3.south);
}
\def\aright[slope=#1,height=#2] (#3) to (#4) node #5; {
    \node (#4) at ($(#3)+({-#1}:{#2/cos(90-#1})$) {#5};
    \draw[thick, <-] (#4.north)--(#3.south);
}
\def\binary[slope left=#1,slope right=#2,left height=#3,right height=#4] (#5) to (#6) node #7 and (#8) node #9; {
    \node (#6) at ($(#5)+({180+#1}:{#3/cos(90-(#1))})$) {#7};
    \node (#8) at ($(#5)+({-#2}:{#4/cos(90-(#2))})$) {#9};
    \draw[thick, <-] (#6.north)--(#5.south);
    \draw[thick, <-] (#8.north)--(#5.south);
}
\node (root) at (0,0) {$X_{255}$};
\aleft[slope=30,height=1.5] (root) to (253) node {$X_{253}$};
\aright[slope=30,height=1.5] (root) to (254) node {$X_{254}$};
\aleft[slope=45,height=1.5] (253) to (249) node {$X_{249}$};
\aright[slope=45,height=1.5] (253) to (250) node {$X_{250}$};
\aleft[slope=45,height=1.5] (254) to (251) node {$X_{251}$};
\aright[slope=45,height=1.5] (254) to (252) node {$X_{252}$};

\aleft[slope=60,height=1.5] (249) to (dots1) node {$\cdots$};
\aright[slope=60,height=1.5] (249) to (dots2) node {$\cdots$};

\aleft[slope=60,height=1.5] (250) to (dots3) node {$\cdots$};
\aright[slope=60,height=1.5] (250) to (dots4) node {$\cdots$};
\aleft[slope=60,height=1.5] (251) to (dots3) node {$\cdots$};
\aright[slope=60,height=1.5] (251) to (dots4) node {$\cdots$};
\aleft[slope=60,height=1.5] (252) to (dots3) node {$\cdots$};
\aright[slope=60,height=1.5] (252) to (dots4) node {$\cdots$};
\aleft[slope=60,height=1.5] (dots1) to (129) node {$X_{129}$};
\aright[slope=60,height=1.5] (dots1) to (130) node {$X_{130}$};
\aleft[slope=70,height=1.5] (129) to (1) node {$X_1$};
\aright[slope=70,height=1.5] (129) to (2) node {$X_2$};
\aleft[slope=70,height=1.5] (130) to (3) node {$X_3$};
\aright[slope=70,height=1.5] (130) to (4) node {$X_4$};
\node (dd) at (-2,-6) {$\cdots$};
\node (dd) at (-2,-7.5) {$\cdots$};
\end{tikzpicture}
\caption{The binary compositional tree considered in Scenarios 1 and 2 of the simulation study with $p = 255$ and $q = 128$. }\label{fig:sim1}
\end{figure}

The first scenario (Scenario 1) has a binary compositional tree and leaf effects. The tree structure is shown in Figure~\ref{fig:sim1}, where $p = 255$, $q = 128$ and $\bX_{leaf} = (X_1, X_2, \dots, X_q)$. Letting $n = 120$, we independently generate $\bX_{leaf,i}, i = 1, \dots, n$ by first independently sampling $\widetilde{\bX}_{leaf,i}$ from a multivariate Gaussian distribution with mean $\boldsymbol{0}_q$ and variance $\boldsymbol{\Sigma} = (\sigma_{ij})_{q\times q}$, where $\sigma_{ij} = 0.2^{|i-j|}$ is the $i$-th row $j$-th column entry of $\boldsymbol{\Sigma}$, and then defining $\bX_{leaf,i} = \widetilde{\bX}_{leaf,i}/\boldsymbol{1}_q^\top\widetilde{\bX}_{leaf,i}$ to satisfy the composition condition. For $j > q$, we generate $X_{ij}$ following the definition of compositional tree using $\bX_{leaf,i}$. We define, for $i = 1,\dots, n$,
\begin{displaymath}
Y_i = 3 + X_{i,1} - X_{i,2} + X_{i,129} - X_{i,130} + \varepsilon = 3 + 2X_{i,1} - X_{i,3} - X_{i,4} + \varepsilon,
\end{displaymath}
where $\varepsilon_i$ is an independent sample from $N(0, \sigma^2)$ and $\sigma^2$ is chosen such that $Var(\bX^\top\bbeta) = Var(\varepsilon)$. This model only involves the left bottom corner in the tree shown in Figure~\ref{fig:sim1}. The non-zero conditional deviation effects are $\beta_1^* = \beta_{129}^* =1, \beta_2^* = \beta_{130}^*= -1$ and the non-zero marginal deviation effects are $\alpha_1^* = 2, \alpha_3^* = \alpha_4^* = -1$.

The second scenario (Scenario 2) has a binary compositional tree and stem effects, where the binary compositional tree and $\bX_i, i=1,\dots,n$ is the same as in Scenario 1. For the stem effect, we define
\begin{displaymath}
Y_i = 3 + X_{i,249} - X_{i,250} + X_{i,253} - X_{i,254} + \varepsilon = 3 + 2\sum_{j=1}^{32}X_{ij} - \sum_{j=65}^{128}X_{ij} + \varepsilon_i,
\end{displaymath}
where $\varepsilon_i$ is defined in the same way as in Scenario 1. Unlike Scenario 1, this model only involves the conditional deviation effects from the top part in the tree (nodes near the root), which are $\beta_{249}^* = \beta_{253}^* =1, \beta_{250}^* = \beta_{254}^*= -1$. Furthermore, the marginal deviation effect $\balpha$ is no longer sparse because $\alpha_j^* = 2$ for $j = 1,\dots, 32$ and $\alpha_j^* = -1$ for $j = 65,\dots, 128$.

In the third scenario (Scenario 3), we consider the MRI-motivated compositional tree and leaf effects. The MRI-motivated compositional tree is shown in Figure~\ref{supp-fig:tree-structure}, where $p = 321$ and $q = 236$. For our MRI data example, $n = 819$ and we denote the empirical distribution of $(\widetilde\bX_1, \dots, \widetilde\bX_n)$ by $F_n$, where $\widetilde\bX_i$ contains the fractional brain volumetric data of participant $i$. Let $\bX_i, i = 1, \dots, n$ be independent samples from $F_n$. We model, for $i=1,\dots,n$,
\begin{displaymath}
Y_i = 3 + 3\ X_{i,\ \textrm{SFG-L}} - 2\ X_{i,\ \textrm{SFG-PFC-L}} - X_{i,\ \textrm{SFG-pole-L}} + \varepsilon,
\end{displaymath}
where SFG-L, SFG-PFC-L and SFG-pole-L are all leaf nodes and subregions of the superior frontal gyrus left hemisphere and $\varepsilon$ is as defined in Scenario 1. In this model, we have $\beta_{\textrm{SFG-L}}^* = \alpha_{\textrm{SFG-L}}^* = 3$, $\beta_{\textrm{SFG-PFC-L}}^* = \alpha_{\textrm{SFG-PFC-L}}^* = -2$ and $\beta_{\textrm{SFG-pole-L}}^* = \alpha_{\textrm{SFG-pole-L}}^* = -1$.

In the last scenario (Scenario 4), we consider the MRI-motivated compositional tree again but with stem effects. We use the same compositional tree and $\bX_i, i = 1, \dots, n$ as in Scenario 3. Let
\begin{displaymath}
Y = 3 + X_{i,\ \textrm{Telencephalon-L}} - X_{i, \ \textrm{Telencephalon-R}} + \varepsilon,
\end{displaymath}
where Telencephalon-L and Telencephalon-R represent telencephalon located in the left and right hemisphere respectively and both are children of the ICV. Different from Scenario 3, this model has $\beta_{\textrm{Telencephalon-L}}^* = -1$, $\beta_{\textrm{Telencephalon-R}}^* = 1$ and 200 non-zero entries in $\balpha^*$.

For each of the 4 scenarios, we simulate $m = 1000$ data sets and compare our proposed method with lasso for component selection (abbreviated as CLASSO throughout, \citealp{lin2014variable}) and TASSO \citep{wang2017structured}. For TASSO, we use their default settings to estimate $\balpha^*$ and calculate $\widehat\bbeta_{TASSO}$ by solving equation~(\ref{eq:estimate-beta}). The only difference is that the natural log-transformation is not performed, as described in Section~\ref{sec:model-assumptions}. Since CLASSO is a special case of our proposed method, we calculate $\widehat\bbeta_{CLASSO}$ following the same procedure as our method setting $\eta = 1$. For all three methods, we use AIC or BIC to select the tuning parameters. The following metrics are used to compare their performances: (1) sensitivity, defined as $|\{j: \widehat{\beta}_j \ne 0, {\beta}_j^* \ne 0\}|/|\{j: {\beta}_j^* \ne 0\}|$, (2) specificity, defined as $|\{j: \widehat{\beta}_j = {\beta}_j^* = 0\}|/|\{j: {\beta}_j^* = 0\}|$ and (3) sum squared error (SSE), defined as $||\widehat{\bbeta} - \bbeta^*||_2^2$. For each of the above metrics, we report its average and standard deviation over the $m$ data sets. Since Scenarios 1 and 2 have $n < q$, a small $L_2$-penalty (0.0001) is added when solving the optimization problem~(\ref{eq:estimate-alpha}).

Table~\ref{sim:table1} gives the simulation results for Scenarios 1-4. 
In Scenarios 1 and 3, the true parameter $\balpha^*$ is sparse and all three methods perform well, as expected. 
For our proposed method, the tuning parameter, $\eta$, is near 0.5, indicating regularization terms on both $\balpha$ and $\bbeta$ help penalize. 
The regularization term on $\bbeta$ is not shown in TASSO and CLASSO and hence leads to the slightly better performance of our method.
In Scenarios 2 and 4, since the true parameter, $\balpha^*$, is not sparse, our proposed method outperforms the other two methods on all performance metrics. In such cases, the $L_1$ penalty on $\balpha$ does not help. Hence, TASSO and CLASSO tend to over-penalize (low sensitivity, high specificity) or under-penalize (high sensitivity, low specificity) on $\balpha$, either leading to high SSEs.
In contrast, our proposed method always selects $\eta = 0$ under BIC tuning, implying it only penalizes on differences of conditional deviation effects. 
Across all 4 scenarios, our method has high accuracy. Compared to AIC, BIC tends to perform as well or better mirroring the simulation results of \cite{wang2017structured}. Hence BIC tuning is used in the MRI data application. In Scenarios 1 and 2, since $n$ is smaller than $q$ and $p$, all three methods have larger SSEs compared to Scenarios 3 and 4.
Our proposed method with BIC tuning, however, remains accurate, suggesting its ability to deal with high-dimensional data. 

In the Supplementary Material, we provide an additional simulation study, where noisier data is used to stress-test the method. In particular, $\sigma^2$ is set such that $ Var(\varepsilon) = 10Var(\bX^\top\bbeta)$ while all of the other settings of Scenarios 1-4 are kept constant. In this simulation, all methods perform worse, because of the weaker signal relative to the noise. Our method, however, still outperforms TASSO and CLASSO and the findings described in other simulations still hold. In addition, simulation results for Scenarios 1 and 2 setting $n = 1000$ are provided, which show similarly good relative performance.

\begin{table}[htbp]
\centering
\caption{Simulation results for Scenarios 1-4 comparing our method, TASSO and CLASSO.}\label{sim:table1}
\resizebox{0.7\textheight}{!}{
\begin{tabular}{cccrrrr}
  \hline
 & Method & Tuning & Sensitivity & Specificity & SSE & $\eta$ \\ 
  \hline
 & Our method & AIC & 1(0.02) & 0.04(0.02) & 20.45(4.94) & 0.31(0.37) \\ 
   &  & BIC & 0.97(0.12) & 0.96(0.08) & 0.8(1.13) & 0.49(0.17) \\ 
  Scenario & TASSO & AIC & 1(0.02) & 0.05(0.02) & 18.64(4.58) & - \\ 
  1 &  & BIC & 0.96(0.18) & 0.96(0.08) & 1.21(1.4) & - \\ 
  & CLASSO & AIC & 1(0) & 0.04(0.02) & 19.41(4.81) & - \\ 
   & & BIC & 0.98(0.14) & 0.91(0.11) & 1.22(1.62) & - \\
  \hline
   & Our method & AIC & 1(0) & 0.01(0.01) & 881.89(205.73) & 0.37(0.38) \\ 
   & & BIC & 1(0) & 0.99(0.06) & 2(27.58) & 0(0.03) \\ 
  Sceinaro & TASSO & AIC & 1(0) & 0.01(0.01) & 846.65(197.18) & - \\ 
  2 & & BIC & 0.04(0.19) & 0.96(0.19) & 28.96(127.08) & - \\ 
    & CLASSO & AIC & 1(0) & 0.01(0.01) & 871.57(204.35) & - \\ 
    & & BIC & 0.42(0.48) & 0.89(0.26) & 57.89(181.14) & - \\ 
    \hline
   & Our method & AIC & 1(0) & 0.85(0.1) & 0.77(0.69) & 0.45(0.33) \\ 
   & & BIC & 1(0) & 0.96(0.03) & 0.78(0.57) & 0.4(0.24) \\ 
  Scenario & TASSO & AIC & 1(0) & 0.92(0.05) & 1.68(0.44) & - \\ 
  3 & & BIC & 1(0) & 0.97(0.04) & 1.58(0.77) & - \\ 
  & CLASSO & AIC & 1(0) & 0.79(0.07) & 2.41(0.4) & - \\ 
  & & BIC & 1(0) & 0.9(0.04) & 2.49(0.4) & - \\ 
  \hline
   & Our method & AIC & 1(0) & 0.93(0.09) & 0.9(3.32) & 0(0.01) \\ 
   & & BIC & 1(0) & 0.98(0.01) & 0.28(0.37) & 0(0) \\ 
  Scenario & TASSO & AIC & 0.92(0.2) & 0.35(0.11) & 34.92(22.27) & - \\ 
  4 & & BIC & 0.48(0.46) & 0.84(0.12) & 13.12(10.08) & - \\ 
   & CLASSO & AIC & 1(0.02) & 0.25(0.05) & 22.3(3.07) & - \\ 
   & & BIC & 0.99(0.11) & 0.59(0.12) & 14.17(4.72) & - \\ 
   \hline
\end{tabular}}
\end{table}

\section{MRI data application}\label{sec:data-analysis}
We applied our proposed method to the data example introduced in Section~\ref{sec:data-example}. The outcome $Y$ is the composite memory score while $\bX$ is the brain volumes resulting from the five-level brain segmentation. Since the simulation study shows that BIC outperforms AIC on the compositional tree of the data example, we used BIC to tune the hyperparameters and obtained $\widehat\eta = 0.405$. 

Our method identified 77 non-zero marginal deviation effects ($\widehat\balpha$) from the 236 leaf brain regions. 
Because of the composition property, each $\widehat\alpha_j$ can be other brain regions as described in Section~\ref{sec:model-assumptions}.
Table~\ref{tab1:data-application} displays the 10 largest effects, which accounts for 48\% of $||\widehat\balpha||_1$. 
Among the 10 largest effects, Hippo-L represents hippocampus in the left hemisphere, which is a limbic subregion and whose atrophy is well established and studied in the progression of AD \citep{pini2016brain}.
InferiorLV-L is the inferior pars of the left lateral ventricle (LV). Evidence has shown that its enlargement is related to MCI and AD \citep{nestor2008ventricular}. The preponderance at CSF structures highlights progressive atrophy in AD. 
Amyg-R stands for amygdala in the right hemisphere. A recent study \citep{poulin2011amygdala} on this region suggested that  ``the magnitude of amygdala atrophy is comparable to that of the hippocampus in the earliest clinical stages of AD, and is related to global illness severity.''
SOG-L, MOG-L and IOG-L represent the left superior, middle and inferior occipital gyri, respectively, and are identified as a group (same marginal deviation effect) by our method.
Cu-L and LG-L are cuneus and lingual gyrus in the left occipital region respectively and also have the same marginal deviation effect.
Although the occipital subregions have opposite signs of marginal deviation effects, their conditional deviation effects (i.e. $\beta_j$) cancel off when combined, resulting in a positive marginal deviation effect (25.06) of the occipital region on memory.
\cite{holroyd2000occipital} showed that occipital atrophy is associated with visual hallucinations (the most common type of hallucination) in AD. However, less is known about the different roles of occipital subregions in AD. 
SylParieSul-L represents sylvian parietal sulcus in the left hemisphere. To the best of our knowledge, its enlargement is associated with progression of AD \citep{liu2012cortical}, which is contrary to our finding. However, we note that this region is also identified with positive marginal deviation effect by TASSO (43.87), which may suggest a false-positive result of the variable selection methods or a special structure of the data set.
MTG-L stands for the left middle temporal gyrus. Its atrophy has been associated with AD \citep{pini2016brain}. However, it is important to emphasize that these results are exploratory in nature, since the method investigates a large possible collection of potential relationships and we did not pre-register any specific hypotheses.

\begin{table}[htbp]
\centering
\caption{Top 10 regression coefficients ($\alpha_j$) of MRI application.}\label{tab1:data-application}
\begin{tabular}{lr|lr}
  \hline
 ROI & $\alpha_j$ & ROI & $\alpha_j$ \\ 
  \hline
Hippo-L & 518.64 &  IOG-L & 90.32  \\ 
  InferiorLV-R & -261.88 & Cu-L & -70.34 \\ 
Amyg-R & 172.73 & LG-L & -70.34 \\ 
 SOG-L & 90.32 & SylParieSul-L & 69.06 \\
MOG-L & 90.32 & MTG-L & 62.14 \\ 
\hline
\end{tabular}
\end{table}

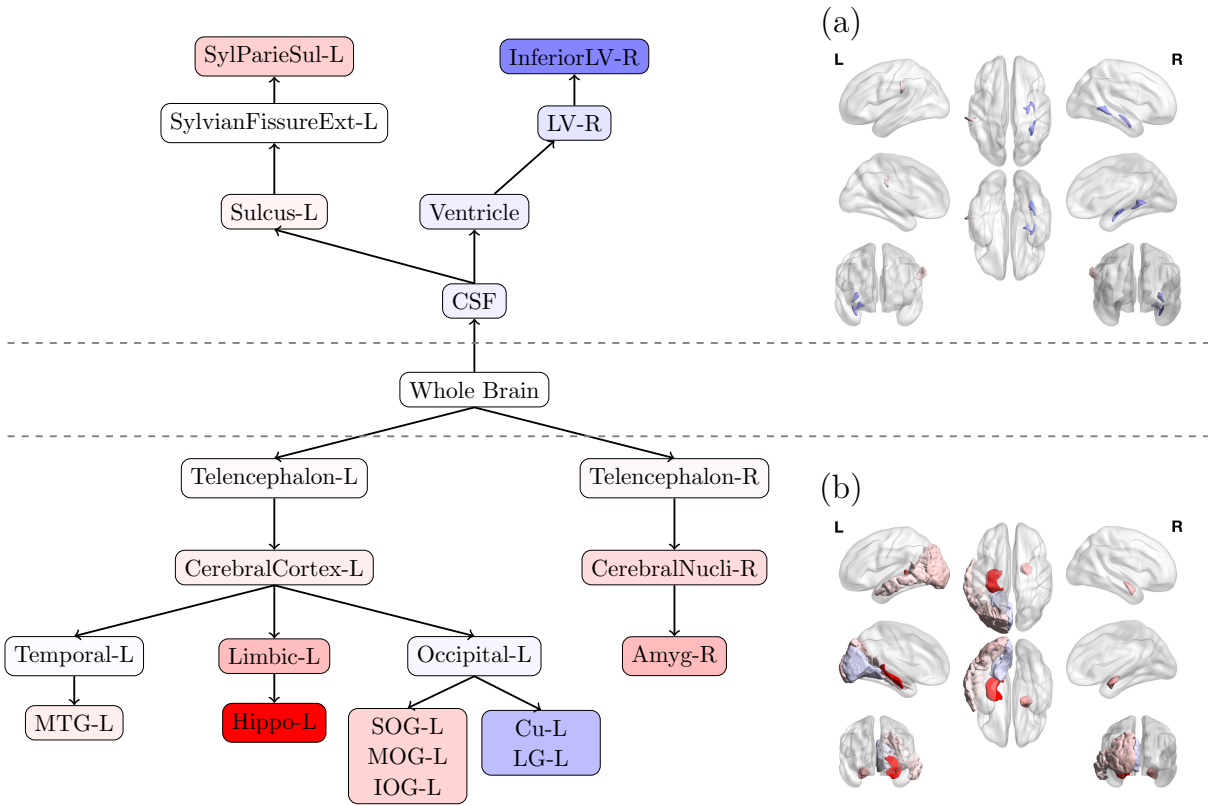
\begin{figure}[htbp]
\centering
\resizebox{\textwidth}{!}{
\begin{tikzpicture}
\tikzstyle{every node}=[font=\footnotesize]
\node[fill={rgb:black,0;white,10},draw,rounded corners] (Intracranial) at (0,0) {Whole Brain};
\node[fill={rgb,255:red,255; green,251; blue,251},draw,rounded corners] (Telencephalon-L) at (-3,-1.33) {Telencephalon-L};
\node[fill={rgb,255:red,255; green,248; blue,248},draw,rounded corners] (Telencephalon-R) at (3,-1.33) {Telencephalon-R};
\node[fill={rgb,255:red,255; green,238; blue,238},draw,rounded corners] (CerebralCortex-L) at (-3,-2.67) {CerebralCortex-L};
\node[fill={rgb,255:red,255; green,219; blue,219},draw,rounded corners] (CerebralNucli-R) at (3,-2.67) {CerebralNucli-R};
\node[fill={rgb,255:red,255; green,188; blue,188},draw,rounded corners] (Amyg-R) at (3,-4) {Amyg-R};
\node[fill={rgb,255:red,251; green,251; blue,255},draw,rounded corners] (Temporal-L) at (-6,-4) {Temporal-L};
\node[fill={rgb,255:red,255; green,189; blue,189},draw,rounded corners] (Limbic-L) at (-3,-4) {Limbic-L};
\node[fill={rgb,255:red,255; green,0; blue,0},draw,rounded corners] (Hippo-L) at (-3,-5) {Hippo-L};
\node[fill={rgb,255:red,246; green,246; blue,255},draw,rounded corners] (Occipital-L) at (0,-4) {Occipital-L};
\node[fill={rgb,255:red,255; green,238; blue,238},draw,rounded corners] (MTG-L) at (-6,-5) {MTG-L};
\node[text width=1.5cm,align=center,fill={rgb,255:red,255; green,212; blue,212},draw,rounded corners] (SOG-L) at (-1,-5.5) {SOG-L\\\vspace{-5pt}MOG-L\\\vspace{-5pt}IOG-L};
\node[text width=1.5cm,align=center,fill={rgb,255:red,190; green,190; blue,255},draw,rounded corners] (LG-L) at (1,-5.3) {Cu-L\\\vspace{-5pt}LG-L};
\draw[thick, ->] (Intracranial.south)--(Telencephalon-L.north);
\draw[thick, ->] (Intracranial.south)--(Telencephalon-R.north);
\draw[thick, ->] (Telencephalon-L) -- (CerebralCortex-L);
\draw[thick, ->] (Telencephalon-R) -- (CerebralNucli-R);
\draw[thick, ->] (CerebralNucli-R) -- (Amyg-R);
\draw[thick, ->] (CerebralCortex-L.south) -- (Temporal-L.north);
\draw[thick, ->] (Temporal-L)--(MTG-L);
\draw[thick, ->] (CerebralCortex-L.south) -- (Occipital-L.north);
\draw[thick, ->] (Occipital-L.south)--(SOG-L.north);
\draw[thick, ->] (Occipital-L.south)--(LG-L.north);
\draw[thick, ->] (CerebralCortex-L.south) -- (Limbic-L.north);
\draw[thick, ->]  (Limbic-L) -- (Hippo-L);

\node[fill={rgb,255:red,240; green,240; blue,255},draw,rounded corners] (CSF) at (0,1.33) {CSF};
\node[fill={rgb,255:red,238; green,238; blue,255},draw,rounded corners] (Ventricle) at (0,2.67) {Ventricle};
\node[fill={rgb,255:red,255; green,245; blue,245},draw,rounded corners] (Sulcus-L) at (-3,2.67) {Sulcus-L};
\node[fill={rgb,255:red,254; green,254; blue,255},draw,rounded corners] (SylvianFissureExt-L) at (-3,4) {SylvianFissureExt-L};
\node[fill={rgb,255:red,229; green,229; blue,255},draw,rounded corners] (LateralVentricle-R) at (1.5,4) {LV-R};
\node[fill={rgb,255:red,133; green,133; blue,255},draw,rounded corners] (InferiorLateralVentricle-R) at (1.5,5) {InferiorLV-R};
\node[fill={rgb,255:red,255; green,208; blue,208},draw,rounded corners] (SylParieSul-L) at (-3,5) {SylParieSul-L};
\draw[thick, ->] (Intracranial)--(CSF);
\draw[thick, ->] (CSF)--(Ventricle);
\draw[thick, ->] (CSF.north)--(Sulcus-L.south);
\draw[thick, ->] (Ventricle)--(LateralVentricle-R);
\draw[thick, ->] (Sulcus-L)--(SylvianFissureExt-L);
\draw[thick, ->] (LateralVentricle-R)--(InferiorLateralVentricle-R.south);
\draw[thick, ->] (SylvianFissureExt-L.north)--(SylParieSul-L.south);

\draw[gray,thick,dashed] (-7,-0.7)--(11,-0.7);
\draw[gray,thick,dashed] (-7,0.7)--(11,0.7);
\node[inner sep=0pt] (TLimage) at (8,-4)
    {\includegraphics[width=.35\textwidth]{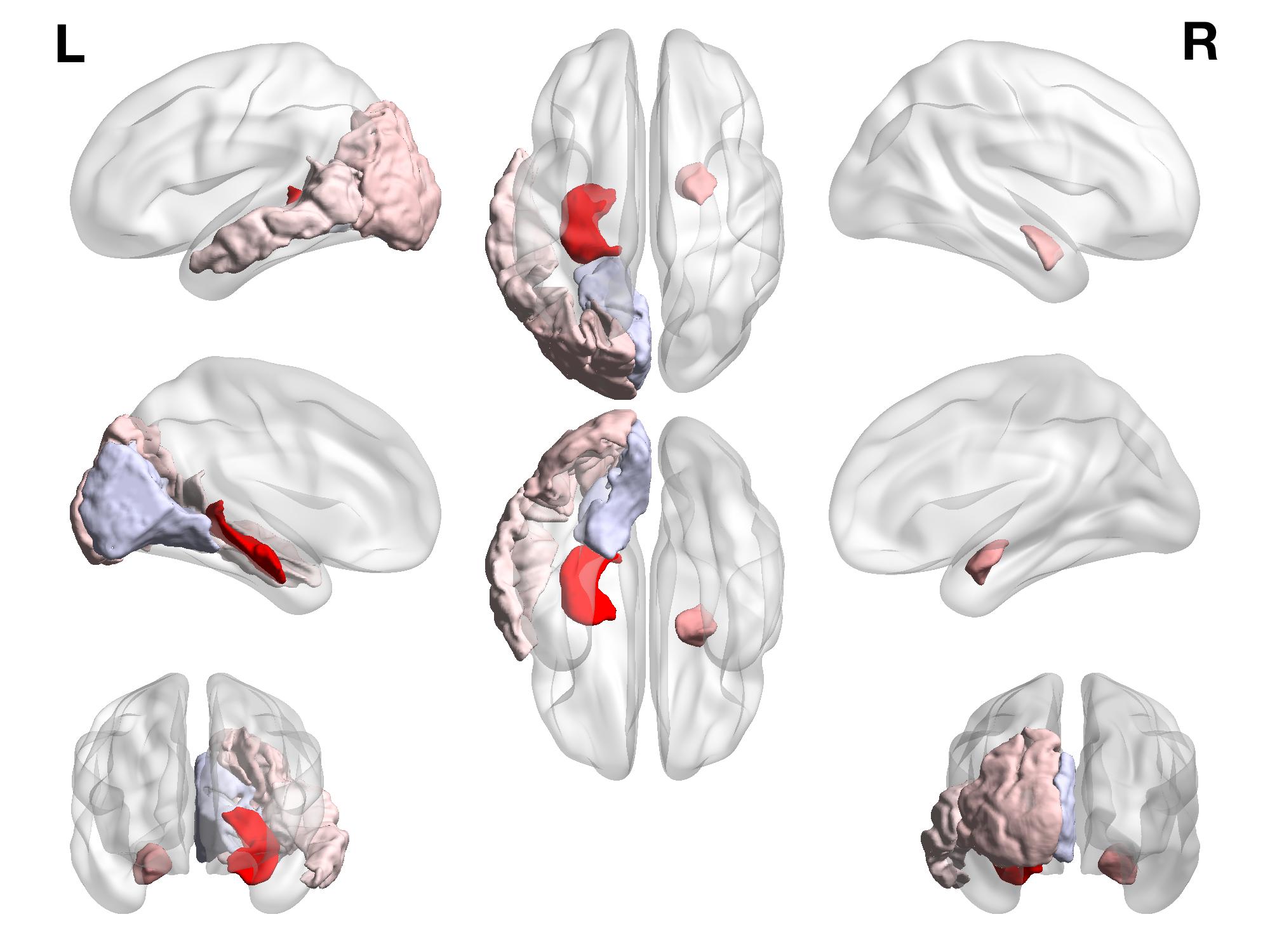}};
\node[inner sep=0pt] (CSFimage) at (8,3)
    {\includegraphics[width=.35\textwidth]{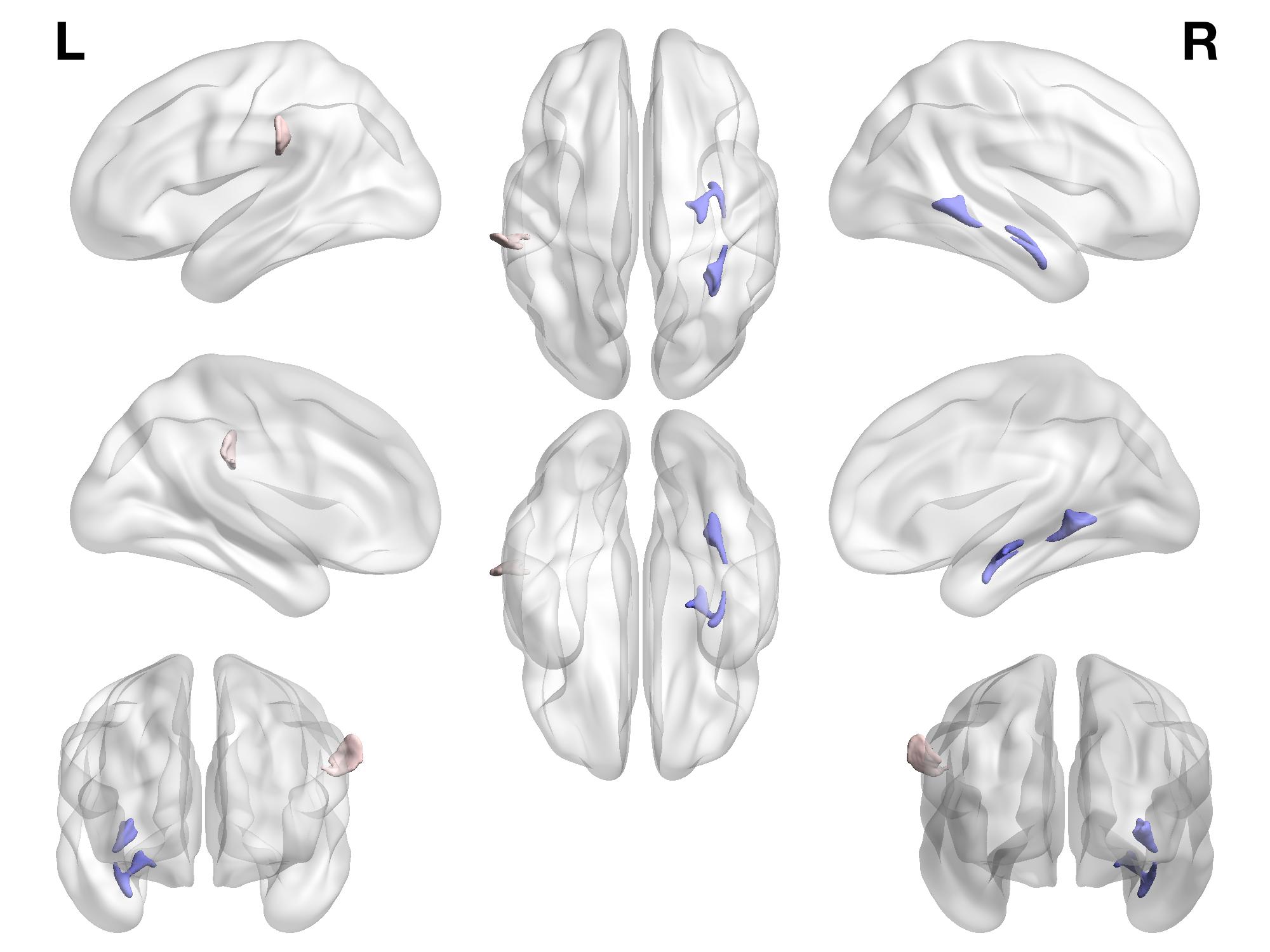}};
\node (a) at (5.5, 5.5) {\large(a)};
\node (b) at (5.5, -1.5) {\large(b)};
\end{tikzpicture}
}

\caption{Conditional deviation effects $\bbeta$ related to the 10 largest marginal deviation effects $\balpha$. Suffix ``-L'' or ``-R'' refers to the left or right hemisphere of the brain, respectively. Red (blue) color represents positive (negative) sign of $\beta$ with darker color indicating a larger value of $|\beta|$. Panels (a) and (b) show the aggregated conditional effects in CSF and telencephalon in 3-dimensional template brain space.  }\label{fig:data-example}
\end{figure}

The 77 marginal deviation effects are aggregations of 109 conditional deviation effects ($\beta$). For the 10 largest marginal deviation effects, we decomposed them into conditional deviation effects using the definition of $\balpha$ (Section~\ref{sec:model-assumptions}) and displayed the results in Figure~\ref{fig:data-example}. All 10 effects are from CSF and telencephalon. The effects from the ventricle are negative and the effects form the limbic region are positive, both of which are consistent with existing scientific findings \citep{pini2016brain, nestor2008ventricular}. Complete results for marginal and conditional deviation effects are given in the Supplementary Material.

In addition to our proposed method, we also ran TASSO and CLASSO with BIC tuning. Twenty non-zero marginal deviation effects and 73 non-zero conditional deviation effects were identified by all three methods, which include brain regions in the ventricles and temporal lobe, although the magnitude of these effects differs substantially among methods.
Especially, the Amyg-R (right hemisphere amygdala) region is only identified by our method.
Compared with our method, TASSO and CLASSO identify fewer marginal effects (40 non-zero entries in $\widehat{\balpha}_{TASSO}$ and 27 non-zero entries in $\widehat{\balpha}_{CLASSO}$), but they have larger BIC (5114 for our proposed method, 5134 for TASSO and 5120 for CLASSO), indicating a larger residual error. 
In addition, our proposed method tends to group effects together, e.g. the left hemisphere occipitial subregions, which can facilitate the interpretation of marginal and conditional deviation effects.
For all three methods, the effects from left and right hemispheres are generally not equal, potentially suggesting a laterally asymmetric correlation between volume and memory.

\section{Discussion}\label{sec:discussion}
The linear model in Section~\ref{sec:model-assumptions} also allows for including additional covariates, in addition to covariates associated with the compositional tree. However, when interaction terms are added, 
the linear constraints~(\ref{eq:beta-constraint}) can only handle interactions between additional covariates and the whole compositional tree.

In our method, for estimating $\balpha^*$, we assume that no components of the leaf nodes are linear combinations of the others such that $\widetilde\balpha^*$ is identifiable. This assumption generally holds if no further linear constraints are made on the leaf nodes. When this assumption is not true, one can add a small $L_2$-penalty to the right side of equation~(\ref{eq:estimate-alpha}) and run the model, otherwise unmodified. In this case, point estimates of $\balpha^*$ and $\bbeta^*$ may be biased because of the $L_2$-penalty.

Our proposed method  also assumes that the outcome is continuous. If the outcome is binary or a count, then relatively minor modifications could use generalized linear models.  However, since the loss function is no longer linear, how to consistently estimate $\balpha^*$ with generalized lasso penalty remains future research.

\section*{Acknowledgements}
Data collection and sharing for this project was funded by the Alzheimer's Disease Neuroimaging Initiative
(ADNI) (National Institutes of Health Grant U01 AG024904) and DOD ADNI (Department of Defense award
number W81XWH-12-2-0012). ADNI is funded by the National Institute on Aging, the National Institute of
Biomedical Imaging and Bioengineering, and through generous contributions from the following: AbbVie,
Alzheimer’s Association; Alzheimer’s Drug Discovery Foundation; Araclon Biotech; BioClinica, Inc.; Biogen;
Bristol-Myers Squibb Company; CereSpir, Inc.; Cogstate; Eisai Inc.; Elan Pharmaceuticals, Inc.; Eli Lilly and
Company; EuroImmun; F. Hoffmann-La Roche Ltd and its affiliated company Genentech, Inc.; Fujirebio; GE
Healthcare; IXICO Ltd.; Janssen Alzheimer Immunotherapy Research \& Development, LLC.; Johnson \&
Johnson Pharmaceutical Research \& Development LLC.; Lumosity; Lundbeck; Merck \& Co., Inc.; Meso
Scale Diagnostics, LLC.; NeuroRx Research; Neurotrack Technologies; Novartis Pharmaceuticals
Corporation; Pfizer Inc.; Piramal Imaging; Servier; Takeda Pharmaceutical Company; and Transition
Therapeutics. The Canadian Institutes of Health Research is providing funds to support ADNI clinical sites
in Canada. Private sector contributions are facilitated by the Foundation for the National Institutes of Health
(www.fnih.org). The grantee organization is the Northern California Institute for Research and Education,
and the study is coordinated by the Alzheimer’s Therapeutic Research Institute at the University of Southern
California. ADNI data are disseminated by the Laboratory for Neuro Imaging at the University of Southern
California.

\bibliographystyle{apalike}
\bibliography{references}

\end{document}


\def\spacingset#1{\renewcommand{\baselinestretch}%
{#1}\small\normalsize}

\title{\bf Supplementary Material to Regularized regression on compositional trees with application to MRI analysis}


\newcommand*{\affaddr}[1]{#1} 
\newcommand*{\affmark}[1][*]{\textsuperscript{#1}}
\newcommand*{\email}[1]{\texttt{#1}}
\author{%
    Bingkai Wang\affmark[1], Brian S. Caffo\affmark[1], Xi Luo\affmark[2], Chin-Fu Liu\affmark[3], Andreia V. Faria\affmark[4],    Michael I. Miller\affmark[3], Yi Zhao\affmark[5],
    and for the Alzheimer's Disease Neuroimaging Initiative\footnote{Data used in preparation of this article were obtained from the Alzheimer's Disease Neuroimaging Initiative (ADNI) database (\protect\url{adni.loni.usc.edu}). As such, the investigators within the ADNI contributed to the design and implementation of ADNI and/or provided data but did not participate in analysis or writing of this report. A complete list of ADNI investigators can be found at: \protect\url{http://adni.loni.usc.edu/wp-content/uploads/how_to_apply/ADNI_Acknowledgement_List.pdf}} \\
    \affaddr{\affmark[1]\small Department of Biostatistics, Johns Hopkins Bloomberg School of Public Health} \\
    \affaddr{\affmark[2]\small Department of Biostatistics and Data Science,   The University of Texas 
Health Science Center at Houston} \\
    \affaddr{\affmark[3]\small Center for Imaging Science, Biomedical Engineering, Johns Hopkins University} \\
    \affaddr{\affmark[4]\small Department of Radiology, Johns Hopkins University School of Medicine} \\    
    \affaddr{\affmark[5]\small Department of Biostatistics, Indiana University School of Medicine} \\    
}
\date{\vspace{-5ex}}
\maketitle

\appendix
\counterwithin{figure}{section}
\counterwithin{table}{section}
\counterwithin{equation}{section}
\counterwithin{theorem}{section}
\spacingset{1.5} 

In Section~\ref{supp:proofs}, we give the proofs of Proposition 1 Proposition 2, Theorem 1 and Theorem 2. In Section~\ref{supp:simulation}, we provide additional simulations. In Section~\ref{supp:data-analysis}, we present the complete result of our proposed method in analysis of the MRI data example.

\section{Proofs}\label{supp:proofs}
Let $||\cdot||_1$, $||\cdot||_2$ and $||\cdot||_\infty$ be the $L_1$, $L_2$ and $L_\infty$-norm of the Euclidean space respectively. For any positive integer $q$,
let $\bI_q \in \mathbb{R}^{q\times q}$ be the identity matrix.
Let $\boldsymbol{1}_q, \boldsymbol{0}_q \in \mathbb{R}^q$ be $q$-dimensional column vectors with all entries 1 and all entries 0 respectively. Let $\be_j$ be a column vector in the Euclidean space such that the $j$-th entry is 1 and all others 0. (The dimension of $\be_j$ is defined according to the context.) For a set $\mathcal{S}$, we define $|\mathcal{S}|$ as the carnality of $\mathcal{S}$.

\subsection{Proof of Proposition 1}
\begin{proof}
Let $(\beta_p, \alpha_1, \dots, \alpha_q)$ and $(\overline\beta_p, \overline\alpha_1, \dots, \overline\alpha_q)$ be two sets of parameters that satisfy model (4), i.e.,
\begin{displaymath}
Y = \beta_p + \sum_{j=1}^q \alpha_j X_j + \varepsilon_j = \overline\beta_p + \sum_{j=1}^q \overline\alpha_j X_j + \varepsilon_j.
\end{displaymath}
This implies that $\sum_{j=1}^q (\alpha_j - \overline{\alpha}_j)X_j = \overline\beta_p - \beta_p$. Since $\sum_{j=1}^q X_j = 1$, then $\sum_{j=1}^q (\overline\beta_p -\beta_p)X_j = \overline\beta_p - \beta_p$. Hence
\begin{displaymath}
\sum_{j=1}^q (\alpha_j - \overline{\alpha}_j -\overline\beta_p + \beta_p )X_j = 0.
\end{displaymath}
By the assumption that $(X_1, \dots, X_p)$ are linearly independent, we have $\alpha_j - \overline{\alpha}_j -\overline\beta_p + \beta_p = 0$ for each $j = 1,\dots, q$, which completes the proof.
\end{proof}

Using Proposition 1, we can also prove the statement in Section 3 of the main paper that linear constraints (2) uniquely determines $\bbeta$. Suppose that two vectors $\bbeta$ and $\overline{\bbeta}$ satisfy model (1). Proposition 1 implies that $\beta_j + \sum_{k \in a(j)} \beta_k = \overline{\beta}_j + \sum_{k \in a(j)} \overline{\beta}_k$ for each $j = 1,\dots, q$. Define the parent of node $j$ as $p(j)$. Then for $l \in c(p(j))$, linear constraints (2) implies that $\sum_{l \in c(p(j))} \beta_l = \sum_{l \in c(p(j))} \overline\beta_l = 0$. Furthermore, $a(l) = a(j)$ since $l$ and $j$ have the same parent. We then have the following derivation:
\begin{align*}
    0 &= \sum_{l \in c(p(j))}\{\beta_l + \sum_{k \in a(l)} \beta_k\} -  \sum_{l \in c(p(j))}\{\overline{\beta}_l + \sum_{k \in a(l)} \overline{\beta}_k\} \\
    &= \sum_{l \in c(p(j))}\sum_{k \in a(l)} \{ \beta_k - \overline{\beta}_k\} \\
    &= |c(p(j))| \sum_{k \in a(j)} \{ \beta_k - \overline{\beta}_k\}.
\end{align*}
Since each $p(j)$ has at least two children, then $|c(p(j))| \ge 1$, which implies that $\beta_j - \overline{\beta}_j = \sum_{k \in a(j)} \{ \beta_k - \overline{\beta}_k\} = 0$ for $j = 1,\dots, q$. By repeating this procedure recursively from the leaf of the tree to the root of the tree, we get $\beta_j = \overline{\beta}_j$ for $j = 1,\dots,p$, which indicates that $\bbeta$ is unique. 
\subsection{Proof of Proposition 2}
\begin{proof}

For $P_1(\balpha)$, let $\bD_1 = \bI_q - \frac{1}{q}\boldsymbol{1}_q\boldsymbol{1}_q^\top$. Direct algebra shows that $\bD_1\boldsymbol{1}_q = \boldsymbol{0}_q$ and $P_1(\balpha) = ||\bD_1\balpha||_1$. 

For $P_2(\bbeta)$, we define $\alpha_j = \sum_{k \in a(j)\setminus\{p\} } \beta_k$ for $j = q+1,\dots p$ and $\overline{\balpha} = (\alpha_1, \dots, \alpha_p)$. Then we have, 
\begin{displaymath}
P_2(\bbeta) = \sum_{j = q+1}^p \sum_{s=1}^{|c(j)|-1}|\alpha_{j_s} - \alpha_{j_{s+1}}| = \sum_{j = q+1}^p ||\bM_j \overline{\balpha}||_1 = ||\bM \overline{\balpha}||_1,
\end{displaymath}
where $\bM_j \in \mathbb{R}^{|c(j)| \times p}$ with the $s$-th row of $\bM_j$ being $(\be_{j_s} - \be_{j_{s+1}})^\top, s = 1,\dots, |c(j)|$ and $\bM = (\bM_{j+1}^\top, \dots, \bM_p^\top)^\top \in \mathbb{R}^{(q-1)\times p}$. To prove the desired result, we still need to connect $\overline{\balpha}$ with $\balpha$.

We next show $\overline{\balpha} = \bH \balpha$, where $\bH \in \mathbb{R}^{p\times q}$ satisfies $\bH \boldsymbol{1}_q = \boldsymbol{1}_p$.
Define $\mathcal{S}_1 = \{1, \dots, q\}$, which represent the index set of the leaf nodes. For $l \ge 2$, define $\mathcal{S}_l = \{j: c(j) \subset \mathcal{S}_{l-1} \}$, which represent the index set of parent nodes for all $X_j, j \in \mathcal{S}_{l-1}$. For example, $\mathcal{S}_2$ contains all parents of leaf nodes and $\mathcal{S}_3$ consists of all parents of nodes in $\mathcal{S}_2$. Given the definition of compositional tree, all nodes except the root have one parent.  Since different nodes may have the same parent, then $|\mathcal{S}_l| \le |\mathcal{S}_{l-1}|$. Since we collapse two nodes if one is the only child of the other, then each non-leaf node has more than 1 children, indicating $|\mathcal{S}_l| < |\mathcal{S}_{l-1}|$. For the decreasing sequence $|\mathcal{S}_{1}|, |\mathcal{S}_{2}|, \dots$, let $L$ be the largest integer such that $|\mathcal{S}_L| >0$. Then $S_L = \{p\}$ since there is only one root in the tree and $\bigcup_{l=1}^L S_l = \{1,\dots, p\}$.
For $l = 2, \dots, L$, denote the elements of $S_l$ as $\{l_1,\dots, l_{|S_l|}\}$ and define $\balpha_{\mathcal{S}_l} = (\alpha_{l_1}, \dots, \alpha_{l_{|S_l|}})$. Then, under linear constraints (2) of the main paper, we have $\balpha_{\mathcal{S}_l} = \bH_l \balpha_{\mathcal{S}_{l-1}}$, where $\bH_l \in \mathbb{R}^{|\mathcal{S}_l| \times |\mathcal{S}_{l-1}|}$ with the $k$-th row being $\frac{1}{|c(l_k)|}\sum_{h \in c(l_k)} \be_{h}$. This is because, for any $j > q$, linear constraints (2) imply $\sum_{h \in c(j)} \beta_h = 0$ and hence
\begin{displaymath}
\alpha_j = \sum_{k \in a(j)\setminus\{p\} } \beta_k = \sum_{k \in a(j)\setminus\{p\} } \beta_k + \frac{1}{|c(l_k)|}\sum_{h \in c(j)} \beta_h = \frac{1}{|c(l_k)|} \sum_{h \in c(j)} \left(\beta_h + \sum_{k \in a(j)\setminus\{p\} } \beta_k\right) = \frac{1}{|c(l_k)|} \sum_{h \in c(j)}\alpha_h.
\end{displaymath}
Furthermore, $\bH_l$ has $\bH_l\boldsymbol{1}_{|\mathcal{S}_{l-1}|} = \boldsymbol{1}_{|\mathcal{S}_l|}$ since $\frac{1}{|c(l_k)|}\sum_{h \in c(l_k)} \be_{h}\boldsymbol{1}_{|\mathcal{S}_{l-1}|} = 1$ for each row $k$ of $\bH_l$. For any $j = 1,\dots, p$, since $\bigcup_{l=1}^L S_l = \{1,\dots, p\}$, we can locate $j$ in an $S_{l(j)}$ at the $k(j)$-th location. Then $$\alpha_j = \be_{k(j)}^t\balpha_{\mathcal{S}_{l(j)}} = \be_{k(j)}^t \left(\prod_{l = 1}^{l(j)} \bH_l \right)\balpha_{\mathcal{S}_{l(1)}} = \be_{k(j)}^t\left(\prod_{l = 1}^{l(j)} \bH_l\right) \balpha.$$
Letting $\bH \in \mathbb{R}^{p\times q}$ be a matrix with $j$-th row being $ \be_{k(j)}^t\prod_{l = 1}^{l(j)} \bH_l$, then $\overline{\balpha} = \bH \balpha$ and $\bH \boldsymbol{1}_q = \boldsymbol{1}_p$, since $ \be_{k(j)}^t\prod_{l = 1}^{l(j)} \bH_l \boldsymbol{1}_q = 1$. 

Finally, $P_2(\bbeta) = ||\bM\overline{\balpha}||_1 = ||\bM\bH\balpha||_1$. Since $\bM$ satisfies $\bM\boldsymbol{1}_p = \boldsymbol{0}_{q-1}$ by its definition, then $\bM\bH\boldsymbol{1}_q = \boldsymbol{0}_{q-1}$. Defining $\bD(\eta) = \left(\begin{array}{c}
    \eta \bD_1  \\
    (1-\eta) \bM \bH
\end{array}\right)$, then $\bD\boldsymbol{1}_q = \boldsymbol{0}_{2q-1}$ and
\begin{displaymath}
P(\balpha, \bbeta, \eta) = \eta P_1(\balpha) + (1-\eta) P_2(\bbeta) = ||\eta\bD_1 \balpha||_1 + ||(1-\eta)\bM\bH\balpha||_1 = ||\bD(\eta)\balpha||_1,
\end{displaymath}
which completes the proof.
\end{proof}
\subsection{Proof of Theorem 1}
Recall that $\mathcal{S}(\balpha)$ is the support of $\bD(\eta)\balpha$ and $\mathcal{M} = \{\balpha: S(\balpha) \subset S(\balpha^*)\}$. We define $\bD_*(\eta) \in \mathbb{R}^{|\mathcal{S}(\balpha^*)|\times q}$ as the rows of $\bD(\eta)$ in $\mathcal{S}(\balpha^*)$ and $\bD_{*,c}(\eta) \in \mathbb{R}^{(2q-1-|\mathcal{S}(\balpha^*)|)\times q}$ as the rows of $\bD(\eta)$ not in $\mathcal{S}(\balpha^*)$.
The loss function of the generalized lasso problem (5) of the main paper is denote as $\ell(\balpha) = \sum_{i=1}^n||Y_i - \balpha^\top \bX_{leaf, i}||_2^2$ and its second derivative is $\nabla^2\ell(\balpha) = \sum_{i=1}^n\bX_{leaf, i}\bX_{leaf, i}^\top$.
We first introduce restricted strong convexity (RSC) and irrepresentability as below.
\begin{description}
  \item[Assumption 1] (Restricted strong convexity, RSC) Let $\mathcal{C}\subset\mathbb{R}^{q}$ be a known convex set containing $\balpha^{*}$. The loss function $\ell(\balpha)$ is RSC on $\mathcal{C}\cap\mathcal{M}$ when
  \begin{equation}
    \btheta^\top \nabla^{2}\ell(\balpha)\btheta\geq m\|\btheta\|_{2}^{2}, ~\balpha\in\mathcal{C}\cap\mathcal{M}, ~\btheta\in(\mathcal{C}\cap\mathcal{M})-(\mathcal{C}\cap\mathcal{M}),
  \end{equation}
  \begin{equation}
    \|\nabla^{2}\ell(\balpha)-\nabla^{2}\ell(\balpha^*)\|_{2}\leq L\|\balpha-\balpha^{*}\|_{2}, ~\bbeta\in\mathcal{C},
  \end{equation}
  for some $m>0$ and $L<\infty$.
  \item[Assumption 2](Irrepresentability) For some $\tau\in(0,1)$,
  \begin{equation}
    \|\bD_{*,c}\bX_{leaf}^\top(\bD_{*}\bX_{leaf}^\top)^{-}\mathrm{sign}\{(\bD(\eta)\balpha^{*})_{\mathcal{S(\balpha^*)}}\}\|_{\infty}\leq 1-\tau,
  \end{equation}
  where $(\bD(\eta)\balpha^{*})_{\mathcal{S(\balpha^*)}}\in\mathbb{R}^{|\mathcal{S(\balpha^*)}|}$ takes the non-zero elements of $\bD(\eta)\balpha^{*}$, $(\bD\bbeta^{*})\in\mathbb{R}^{|\mathcal{S}|}$, and $\bA^{-}$ is the \textit{Moore-Penrose pseudoinverse} of a matrix $\bA\in\mathbb{R}^{q\times q}$ and $\mathrm{sign}(\cdot)$ is the sign function.
\end{description}

We next prove Theorem 1 of the main paper.
\begin{proof}
Given our assumptions, Theorem 1 is equivalent to Corollary 4.2 of \cite{lee2015model} and the proof can be found accordingly. The only unspecified quantities are $C_1$ and $C_2$, which we give below. These quantities are the same as those given in \cite{lee2015model}.

We define
\begin{align*}
    \kappa_{\rho} &= \sup_{\btheta}\{||\bD(\eta)\btheta||_1: \btheta \in B_2 \cap \mathcal{M}\}, \\
    \kappa_{\varrho} &= \sup_{\btheta}\{||\btheta||_1: \btheta \in B_2 \cap \mathcal{M}\}, \\ 
    \kappa_{IC} &= \sup_{||\bz||_1^* \le 1}|| \bD_{c,*}\bX_{leaf}^\top (\bD_{*}\bX_{leaf}^\top)^{-} \bz_{\mathcal{S}(\balpha^*)} - \bz_{\mathcal{S}(\balpha^*)}^c ||_{\infty},
\end{align*}
where $B_2 = \{\bz \in \mathbb{R}^q: ||\bz||_2 \le 1\}$ is the unit ball in $\mathbb{R}^q$, $||\bz||_1^* = \sup_{||\boldsymbol{y}||_1\le 1}\boldsymbol{y}^\top \bz$ is the dual norm of $||\cdot||_1$, $\bz_{\mathcal{S}(\balpha^*)} \in \mathbb{R}^{|\mathcal{S(\balpha^*)}|}$ takes the elements of $\bz$ in $\mathcal{S(\balpha^*)}$ and $\bz_{\mathcal{S}(\balpha^*)} \in \mathbb{R}^{|\mathcal{S(\balpha^*)}|}$ takes the elements of $\bz$ not in $\mathcal{S(\balpha^*)}$. Then we have
\begin{displaymath}
C_1 = \frac{8\kappa_{IC}}{\tau}, C_2 = \frac{4}{2q-1}\left(\kappa_{\varrho} + \frac{4\kappa_{IC}}{\tau}\kappa_{\rho}\right)
\end{displaymath}
following Corollary 4.2 of \cite{lee2015model}.
\end{proof}
\subsection{Proof of Theorem 2}
\begin{proof}
In this proof, we first show that $\bQ$ is invertible and then prove the consistency of $\widehat\bbeta$. We define $\mathcal{S}_1 = \{1, \dots, q\}$, which represent the index set of the leaf nodes. For $l \ge 2$, define $\mathcal{S}_l = \{j: c(j) \subset \mathcal{S}_{l-1} \}$, which represent the index set of parent nodes for all $X_j, j \in \mathcal{S}_{l-1}$. Following the statements in the proof of Proposition 1, let $L$ be the largest integer such that $|\mathcal{S}_L| > 0$ and we have $L$ is finite, $\mathcal{S}_L = \{p\}$ and $\bigcup_{l=1}^L S_l = \{1,\dots, p\}$. For each $j = 1, \dots, p$, we define $s(j)$ as the index set of siblings of $X_j$, i.e., $X_j$ and $X_t, t \in s(j)$ have the same parent, and $o(j)$ as the offspring of $X_j$, i.e., $o(j) = \{t: j \in a(j)\}$.

In order to show that $\bQ$ is invertible, we construct $\bz_j  \in \mathbb{R}^{p }$ such that $\bz_j^\top \bQ = \be_j^\top$ for each $j = 1,\dots, p$ and $\be_j \in \mathbb{R}^p$. We prove this statement by induction on $j \in S_l, l = 1,\dots, L$. Letting $l = 1$, for $j \in \mathcal{S}_1$, since $\alpha_j = \beta_j + \sum_{k \in a(j)} \beta_k = \beta_j + \sum_{k \in a(j)} \be_k^\top \bbeta$, there exists a row $m(j)$ of $\bQ$ such that $\be_{m(j)}^\top \bQ = \be_j^\top + \sum_{k \in a(j)} \be_k^\top$ and $\balpha_j = e_{m(j)}^\top \bQ \bbeta$. Similarly, for each $t \in s(j)$, we can also find $m(t)$ such that $e_{m(t)}^\top \bQ = \be_t^\top + \sum_{k \in a(t)} \be_k^\top$. Furthermore, since we have constraint $\beta_j + \sum_{t \in s(j)} \beta_t = 0$, then there exists a row $n(j)$ of $\bQ$ such that $\be_{n(j)}^\top\bQ = \be_j^\top + \sum_{t \in s(j)} \be_t^\top$. Since $a(j) = a(t)$ for $t \in s(j)$, then
\begin{displaymath}
\be_{m(j)}^\top\bQ - \frac{1}{|s(j)|+ 1}\left(
\be_{m(j)}^\top\bQ + \sum_{t \in s(j)}\be_{m(t)}^\top\bQ - \be_{n(j)}^\top \bQ\right) = \be_j^\top. 
\end{displaymath}
Hence $\bz_j = \frac{|s(j)|}{|s(j)|+ 1}\be_{m(j)} - \frac{1}{|s(j)|+ 1}\sum_{t \in s(j)}\be_{m(t)} + \frac{1}{|s(j)|+ 1}\be_{n(j)}$ satisfies $\bz_j^\top\bQ = \be_j^\top$ for $j \in \mathcal{S}_1$. Next we assume that there exists $\bz_j  \in \mathbb{R}^{p}$ such that $\bz_j^\top\bQ = \be_j^\top$ for $j \in \mathcal{S}_1\cup\dots\cup \mathcal{S}_{l-1}$ and show that there exists $\bz_j  \in \mathbb{R}^{p}$ such that $\bz_j^\top\bQ = \be_j^\top$ for $j \in \mathcal{S}_l$. For any $j \in \mathcal{S}_l$ and an $r \in o(j)$, since $\alpha_r = \beta_r + \sum_{t \in a(r)\cap o(j)}\beta_{t} + \beta_j + \sum_{t \in a(j)} \beta_t$, then there exists a row $m(r)$ of $\bQ$ such that $\be_{m(r)}^\top \bQ = \be_r^\top + \sum_{t \in a(r)\cap o(j)}\be_t^\top + \be_j^\top + \sum_{t \in a(j)}\be_t^\top$. Since $o(j) \subset \mathcal{S}_1\cup\dots\cup \mathcal{S}_{l-1}$, by assumption, for $t \in o(j)$, we can find $\bz_t$ such that $\bz_t^\top\bQ = \be_t^\top$. Hence, letting $\widetilde\bz_{j} = \be_{m(r)}-\bz_r - \sum_{t \in a(r)\cap o(j)} \bz_t$, we have $\widetilde\bz_{j}^\top\bQ = \be_j^\top + \sum_{t \in a(j)}\be_t^\top$. Similarly, for $t \in s(j)$, we can found $\widetilde{\bz}_{t}$ such that $\widetilde\bz_{t}^\top\bQ = \be_{t}^\top + \sum_{k \in a(t)}\be_k^\top$. Furthermore, since we have constraint $\beta_j + \sum_{t \in s(j)} \beta_t = 0$, then there exists a row $n(j)$ of $\bQ$ such that $\be_{n(j)}^\top\bQ = \be_j^\top + \sum_{t \in s(j)} \be_t^\top$. Since $a(j) = a(t)$ for $t \in s(j)$, then
\begin{displaymath}
\widetilde{\bz}_{j}\bQ - \frac{1}{|s(j)|+ 1}\left(
\widetilde{\bz}_{j}\bQ + \sum_{t \in s(j)}\widetilde{\bz}_{t}^\top\bQ - \be_{n(j)}^\top \bQ\right) = \be_j^\top. 
\end{displaymath}
Hence $\bz_j = \frac{|s(j)|}{|s(j)|+ 1}\widetilde{\bz}_{j} - \frac{1}{|s(j)|+ 1}\sum_{t \in s(j)}\widetilde{\bz}_{t} + \frac{1}{|s(j)|+ 1}\be_{n(j)}$ satisfies $\bz_j^\top\bQ = \be_j^\top$ for $j \in \mathcal{S}_l$, which proves that $\bQ$ is invertible by induction.

For proving the consistency of $\widehat\bbeta$, we have
\begin{displaymath}
||\widehat\bbeta - \bbeta^*||_2 \le ||\bQ^{-1}||_2 ||\bQ(\widehat\bbeta - \bbeta^*)||_2 = ||\bQ^{-1}||_2 ||((\widehat\balpha - \balpha^*)^\top, \boldsymbol{0}_{p-q}^\top)^\top||_2 = ||\bQ^{-1}||_2 ||\widehat\balpha - \balpha^*||_2.
\end{displaymath}
Under the same assumption of Theorem 1, we have $||\widehat\balpha - \balpha^*||_2 \le C_2 \sigma\sqrt{\frac{\log q}{n}}$. Hence $||\widehat\bbeta - \bbeta^*||_2 \le ||\bQ^{-1}||_2C_2 \sigma\sqrt{\frac{\log q}{n}}$. For the model selection consistency, since $\bQ_1\widehat\bbeta = \widehat\balpha$, then $\bQ_1\widehat\bbeta \in \mathcal{M}$ is implied by Theorem 1.
\end{proof}

\section{Additional simulations}\label{supp:simulation}
The simulation results for for Scenarios 1 and 2 setting $n = 1000$ are given in Table~\ref{supp:sim-table1}. We provide results for the additional simulation study (Scenarios 1-4 setting $ Var(\varepsilon) = 10Var(\bX^\top\bbeta)$) in Table~\ref{supp:sim-table2}.
\begin{table}[htbp]
\centering
\caption{Simulation results for Scenarios 1 and 2 comparing our method, TASSO and CLASSO setting $n = 1000$ and $ Var(\varepsilon) = Var(\bX^\top\bbeta)$.}\label{supp:sim-table1}
\resizebox{0.7\textheight}{!}{
\begin{tabular}{cccrrrr}
  \hline
 & Method & Tuning & Sensitivity & Specificity & SSE & $\eta$ \\ 
  \hline
 & our method & AIC & 1(0) & 0.89(0.10) & 0.05(0.03) & 0.55(0.25) \\ 
   &   & BIC & 1(0) & 0.99(0.01) & 0.05(0.03) & 0.54(0.25) \\ 
  Scenario & TASSO & AIC & 1(0) & 0.88(0.08) & 0.09(0.05) & - \\ 
  1 &   & BIC & 1(0) & 0.98(0.01) & 0.09(0.05) & - \\ 
    & CLASSO & AIC & 1(0) & 0.84(0.11) & 0.07(0.04) & - \\ 
    &   & BIC & 1(0) & 0.97(0.02) & 0.07(0.04) & - \\ \hline
    & our method & AIC & 1(0) & 0.95(0.07) & 0.36(0.94) & 0.01(0.03) \\ 
    &   & BIC & 1(0) & 1(0) & 0.03(0.03) & 0.01(0.02) \\ 
  Scenario & TASSO & AIC & 1(0) & 0.03(0.02) & 25.59(4.11) & - \\ 
  2 &   & BIC & 0.32(0.47) & 0.69(0.45) & 10.68(9.89) & - \\ 
    & CLASSO & AIC & 1(0) & 0.05(0.02) & 20.61(3.54) & - \\ 
    &   & BIC & 1(0) & 0.29(0.10) & 12.91(2.72) & - \\ \hline
\end{tabular}}
\end{table}

\begin{table}[htbp]
\centering
\caption{Simulation results for Scenarios 1-4 comparing our proposed method, TASSO and CLASSO setting $ Var(\varepsilon) = 10Var(\bX^\top\bbeta)$ .}\label{supp:sim-table2}
\resizebox{0.7\textheight}{!}{
\begin{tabular}{cccrrrr}
  \hline
 & method & tuning & sensitivity & specifity & SSE & parameter \\ 
  \hline
 & our method & AIC & 0.99(0.05) & 0.02(0.01) & 195.33(47.13) & 0.35(0.38) \\ 
   &  & BIC & 0.36(0.43) & 0.97(0.07) & 3.53(7.28) & 0.52(0.31) \\ 
  Scenario & TASSO & AIC & 1(0.04) & 0.02(0.01) & 184.28(44.33) & - \\ 
  1 &  & BIC & 0.14(0.31) & 1(0.03) & 3.88(4.87) & - \\ 
    & CLASSO & AIC & 1(0.05) & 0.01(0.01) & 191.71(46.21) & - \\ 
    &  & BIC & 0.27(0.43) & 0.98(0.05) & 3.74(7.05) & - \\ 
    \hline
   & our method & AIC & 0.99(0.06) & 0.01(0.01) & 8688.23(2109.03) & 0.32(0.37) \\ 
   &  & BIC & 0.66(0.39) & 0.99(0.03) & 4.44(7.5) & 0.18(0.3) \\ 
  Scenario & TASSO & AIC & 0.96(0.14) & 0(0) & 8256.03(2023.24) & - \\ 
  2 &  & BIC & 0(0) & 1(0.01) & 9.7(22.51) & - \\ 
   & CLASSO & AIC & 0.99(0.05) & 0(0) & 8579.73(2065.31) & - \\ 
   & & BIC & 0.1(0.29) & 0.99(0.04) & 10.95(145.11) & - \\ 
  \hline
   & our method & AIC & 1(0.01) & 0.82(0.12) & 3.24(3.59) & 0.41(0.35) \\ 
   & & BIC & 0.99(0.11) & 0.96(0.04) & 3.45(2.1) & 0.39(0.24) \\ 
  Scenario & TASSO & AIC &  0.95(0.12) & 0.94(0.05) & 3.64(14.03) & - \\ 
  3 &   & BIC & 0.84(0.2) & 0.98(0.01) & 4.52(2.2) & - \\ 
   & CLASSO & AIC & 1(0.02) & 0.8(0.09) & 4.54(1.48)  & - \\ 
   &   & BIC & 0.98(0.13) & 0.93(0.03) & 7.03(1.96) & - \\ 
   \hline
   & our method & AIC & 1(0) & 0.92(0.09) & 6.23(22.15) & 0.01(0.06) \\ 
   & & BIC & 1(0.05) & 0.98(0.01) & 0.6(0.44) & 0(0.04) \\ 
  Scenario & TASSO & AIC & 0.28(0.36) & 0.85(0.11) & 17.28(14.9) & - \\ 
  4 &  & BIC & 0.01(0.09) & 0.99(0.03) & 2.39(1.73) & - \\ 
   & CLASSO & AIC & 0.95(0.2) & 0.6(0.14) & 20.76(10.69) & - \\ 
   &  & BIC & 0.24(0.38) & 0.96(0.06) & 3.98(4.07) & - \\ 
   \hline
\end{tabular}}
\end{table}

\section{Additional results of MRI data application} \label{supp:data-analysis}
Complete marginal and conditional deviation effects are given in Tables~\ref{supp-tab:alpha} and \ref{supp-tab:beta} respectively.

\begin{figure}[htbp]
    \centering
    \includegraphics[width=0.001\textwidth]{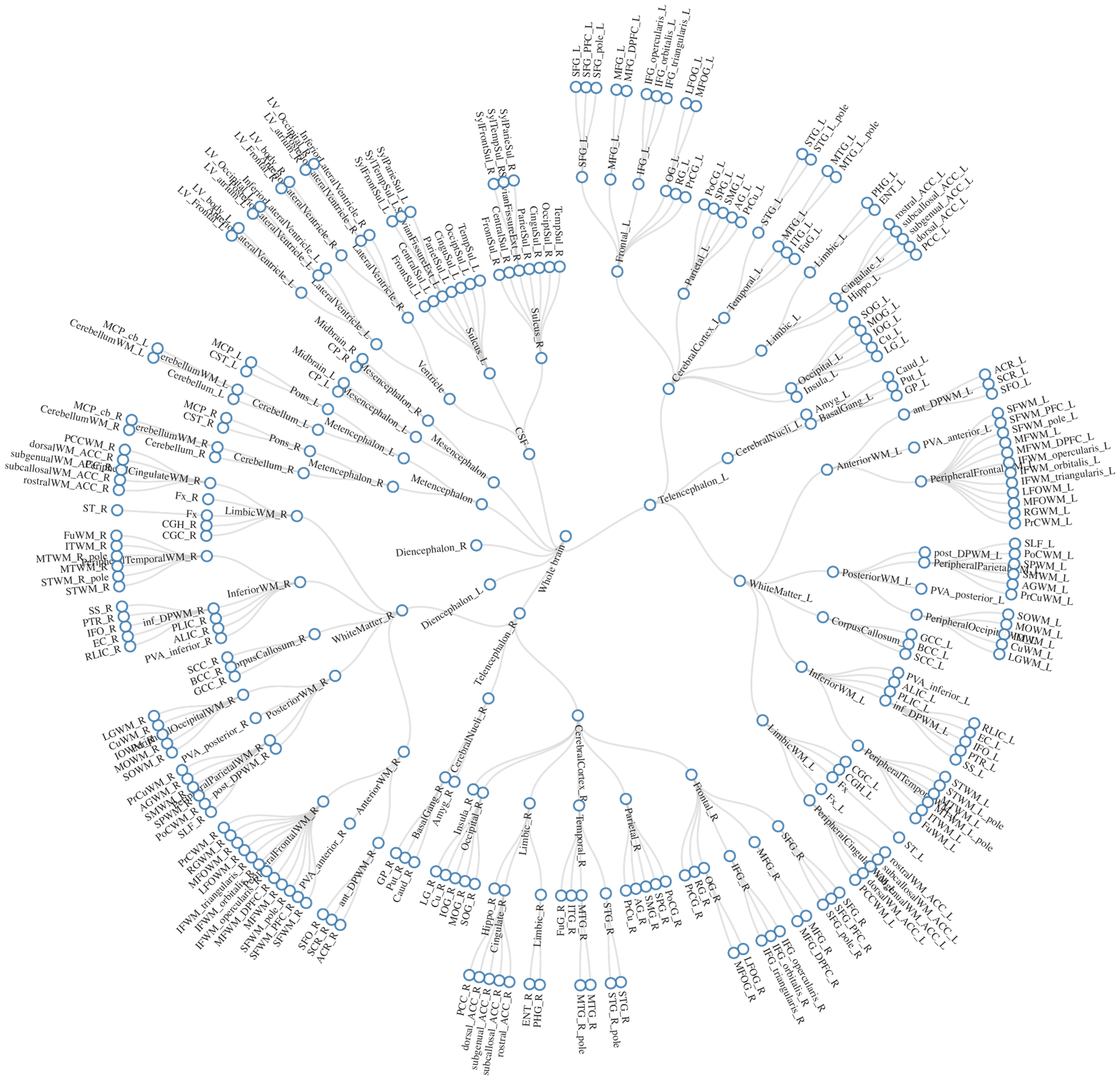}
\end{figure}

\begin{table}[htbp]
\centering
\caption{Non-zero marginal deviation effect ($\widehat\balpha$) by our proposed method.}\label{supp-tab:alpha}
\resizebox{0.88\textwidth}{!}{
\begin{tabular}{lrlrlr}
  \hline
names & alpha & names & alpha & names & alpha \\ 
  \hline
Hippo-L & 518.64 & PCC-R & 35.36 & IOG-R & -18.05 \\ 
  InferiorLateralVentricle-R & -261.88 & PoCG-L & -32.54 & Cu-R & -18.05 \\ 
  Amyg-R & 172.73 & FrontSul-R & -27.95 & LG-R & -18.05 \\ 
  SOG-L & 90.32 & Caud-R & -27.63 & Insula-R & -18.05 \\ 
  MOG-L & 90.32 & Put-R & -27.63 & FuG-L & 17.41 \\ 
  IOG-L & 90.32 & GP-R & -27.63 & Cerebellum-R & -11.48 \\ 
  Cu-L & -70.34 & Insula-L & 26.06 & LV-Frontal-R & 10.68 \\ 
  LG-L & -70.34 & SPG-R & -24.26 & LV-body-R & 10.68 \\ 
  SylParieSul-L & 69.06 & SMG-R & -24.26 & LV-atrium-R & 10.68 \\ 
  MTG-L & 62.14 & AG-R & -24.26 & LV-Occipital-R & 10.68 \\ 
  PoCG-R & -57.27 & PrCu-R & -24.26 & FrontSul-L & 10.58 \\ 
  PHG-L & -52.56 & LV-Frontal-L & -23.56 & MTG-L-pole & 10.27 \\ 
  ENT-L & -52.56 & SPG-L & 22.75 & ITG-R & 10.13 \\ 
  rostral-ACC-L & -52.56 & SMG-L & 22.75 & FuG-R & 10.13 \\ 
  subcallosal-ACC-L & -52.56 & AG-L & 22.75 & PHG-R & -7.40 \\ 
  subgenual-ACC-L & -52.56 & PrCu-L & 22.75 & ENT-R & -7.40 \\ 
  dorsal-ACC-L & -52.56 & MTG-R & 22.35 & CentralSul-L & -5.42 \\ 
  PCC-L & -52.56 & PrCG-R & -22.16 & LV-body-L & 5.26 \\ 
  STG-R & -48.88 & Diencephalon-L & 18.88 & Cerebellum-L & 3.08 \\ 
  STG-R-pole & -48.88 & Diencephalon-R & 18.88 & MTG-R-pole & -2.09 \\ 
  Hippo-R & -46.08 & rostral-ACC-R & -18.09 & LV-atrium-L & 1.71 \\ 
  STG-L & 44.77 & subcallosal-ACC-R & -18.09 & LV-Occipital-L & 1.71 \\ 
  STG-L-pole & 44.77 & subgenual-ACC-R & -18.09 & InferiorLateralVentricle-L & 1.71 \\ 
  ITG-L & 36.20 & dorsal-ACC-R & -18.09 & ParietSul-L & -1.05 \\ 
  SylFrontSul-L & -36.11 & SOG-R & -18.05 & CinguSul-L & -1.05 \\ 
  SylTempSul-L & -36.11 & MOG-R & -18.05 &  &  \\ 
   \hline
\end{tabular}}
\end{table}

\begin{table}[htbp]
\centering
\caption{Non-zero conditional deviation effect ($\widehat\bbeta$) by our proposed method.}\label{supp-tab:beta}
\resizebox{0.8\textwidth}{!}{
\begin{tabular}{lrlrlr}
  \hline
names & beta & names & beta & names & beta \\ 
  \hline
Hippo\_L.lvl4 & 380.80 & Sulcus\_L.lvl2 & 15.20 & PrCu\_R.lvl4 & 6.60 \\ 
  Limbic\_L.lvl4 & -190.40 & Parietal\_R.lvl3 & -14.94 & Cerebellum\_R.lvl4 & -5.74 \\ 
  Cingulate\_L.lvl4 & -190.40 & ITG\_R.lvl4 & 14.75 & CerebellumWM\_R.lvl4 & 5.74 \\ 
  InferiorLateralVentricle\_R.lvl4 & -181.70 & FuG\_R.lvl4 & 14.75 & CentralSul\_L.lvl3 & -5.70 \\ 
  Amyg\_R.lvl3 & 100.18 & MTG\_R.lvl4 & 14.75 & Temporal\_L.lvl3 & -5.57 \\ 
  BasalGang\_R.lvl3 & -100.18 & LV\_Frontal\_L.lvl5 & -14.41 & Telencephalon\_L.lvl1 & 5.40 \\ 
  Limbic\_L.lvl3 & 98.63 & LV\_body\_L.lvl5 & 14.41 & Limbic\_R.lvl3 & -4.37 \\ 
  Cu\_L.lvl4 & -96.39 & Insula\_L.lvl3 & -13.16 & CentralSul\_R.lvl3 & 3.99 \\ 
  LG\_L.lvl4 & -96.39 & Occipital\_L.lvl3 & -13.16 & ParietSul\_R.lvl3 & 3.99 \\ 
  AnteriorLateralVentricle\_R.lvl4 & 90.85 & CerebralNucli\_L.lvl2 & -13.07 & CinguSul\_R.lvl3 & 3.99 \\ 
  PosteriorLateralVentricle\_R.lvl4 & 90.85 & WhiteMatter\_L.lvl2 & -13.07 & OcciptSul\_R.lvl3 & 3.99 \\ 
  SylParieSul\_L.lvl4 & 70.12 & Limbic\_R.lvl4 & 12.89 & TempSul\_R.lvl3 & 3.99 \\ 
  SOG\_L.lvl4 & 64.26 & Cingulate\_R.lvl4 & 12.89 & SylvianFissureExt\_R.lvl3 & 3.99 \\ 
  MOG\_L.lvl4 & 64.26 & Frontal\_R.lvl3 & 12.23 & RG\_R.lvl4 & 3.69 \\ 
  IOG\_L.lvl4 & 64.26 & MTG\_R.lvl5 & 12.22 & SFG\_R.lvl4 & 3.69 \\ 
  CerebralNucli\_R.lvl2 & 53.68 & MTG\_R\_pole.lvl5 & -12.22 & MFG\_R.lvl4 & 3.69 \\ 
  STG\_R.lvl4 & -44.26 & Temporal\_R.lvl3 & 11.31 & IFG\_R.lvl4 & 3.69 \\ 
  PoCG\_L.lvl4 & -44.23 & Diencephalon\_L.lvl1 & 11.20 & OG\_R.lvl4 & 3.69 \\ 
  PCC\_R.lvl5 & 42.76 & Diencephalon\_R.lvl1 & 11.20 & InferiorLateralVentricle\_L.lvl4 & 3.62 \\ 
  Frontal\_L.lvl3 & -39.22 & Telencephalon\_R.lvl1 & 11.20 & PosteriorLateralVentricle\_L.lvl4 & 3.62 \\ 
  LateralVentricle\_L.lvl3 & 39.13 & STG\_L.lvl4 & 11.12 & Pons\_R.lvl3 & 2.87 \\ 
  LateralVentricle\_R.lvl3 & -39.13 & SPG\_L.lvl4 & 11.06 & Cerebellum\_R.lvl3 & -2.87 \\ 
  SylFrontSul\_L.lvl4 & -35.06 & SMG\_L.lvl4 & 11.06 & ITG\_L.lvl4 & 2.56 \\ 
  SylTempSul\_L.lvl4 & -35.06 & AG\_L.lvl4 & 11.06 & MTG\_L.lvl4 & 2.56 \\ 
  CerebralCortex\_R.lvl2 & -34.80 & PrCu\_L.lvl4 & 11.06 & Insula\_R.lvl3 & -2.12 \\ 
  Parietal\_L.lvl3 & -27.53 & Sulcus\_R.lvl2 & 10.92 & Occipital\_R.lvl3 & -2.12 \\ 
  PoCG\_R.lvl4 & -26.41 & rostral\_ACC\_R.lvl5 & -10.69 & Metencephalon\_R.lvl2 & -1.82 \\ 
  CerebralCortex\_L.lvl2 & 26.14 & subcallosal\_ACC\_R.lvl5 & -10.69 & Metencephalon\_L.lvl2 & 1.82 \\ 
  Ventricle.lvl2 & -26.13 & subgenual\_ACC\_R.lvl5 & -10.69 & Cerebellum\_L.lvl4 & 1.54 \\ 
  MTG\_L.lvl5 & 25.94 & dorsal\_ACC\_R.lvl5 & -10.69 & CerebellumWM\_L.lvl4 & -1.54 \\ 
  MTG\_L\_pole.lvl5 & -25.94 & FrontSul\_L.lvl3 & 10.29 & ParietSul\_L.lvl3 & -1.34 \\ 
  Hippo\_R.lvl4 & -25.79 & Metencephalon.lvl1 & -8.73 & CinguSul\_L.lvl3 & -1.34 \\ 
  FrontSul\_R.lvl3 & -23.95 & Mesencephalon.lvl1 & -7.68 & SylvianFissureExt\_L.lvl3 & -1.34 \\ 
  CSF.lvl1 & -22.59 & AnteriorLateralVentricle\_L.lvl4 & -7.24 & Pons\_L.lvl3 & -0.77 \\ 
  WhiteMatter\_R.lvl2 & -18.88 & SPG\_R.lvl4 & 6.60 & Cerebellum\_L.lvl3 & 0.77 \\ 
  PrCG\_R.lvl4 & -18.47 & SMG\_R.lvl4 & 6.60 &  \\ 
  FuG\_L.lvl4 & -16.23 & AG\_R.lvl4 & 6.60 &  \\ 
   \hline
\end{tabular}}
\end{table}

\bibliographystyle{apalike}
\bibliography{references}